\documentclass[11pt]{article}
\usepackage[dvips]{graphicx}  

\begin{document}

\title{{\bf `DARK MATTER' AS A QUANTUM FOAM IN-FLOW EFFECT}}

\author{{ \bf Reginald T. Cahill}\\
School of Chemistry, Physics and Earth Sciences\\
 Flinders University \\
 GPO Box 2100, Adelaide 5001, Australia \\
 Reg.Cahill@flinders.edu.au \\ \\ \\  
Published in  {\it Trends in Dark
Matter Research}, \\ ed J. Val Blain,   Nova Science Pub. NY 2005.
\\  arXiv:physics/0405147  May 28, 2004  
}

\setcounter{page}{1}

\date{}

\maketitle

\begin{center} Abstract \end{center}

The galactic `dark matter' effect  is regarded as one of the major problems in
fundamental physics.   Here it  is explained  as a self-interaction dynamical effect of space itself, and so is
not caused by an unknown form of matter.   Because it was  based on Kepler's Laws for
the motion of the planets in the solar system  the Newtonian theory of gravity  was too restricted.
A reformulation and generalisation of the Newtonian theory of gravity in terms of a velocity in-flow
field, representing at a classical level the relative motion of  a quantum-foam substructure to
space,  reveals a key dynamical feature of the phenomenon of gravity, namely the so called `dark
matter' effect, which manifests not only in spiral galaxy rotation curves, but also in the  borehole
$g$ anomaly, globular and galactic black holes, and in ongoing problems in improving the accuracy
with which Newton's gravitational constant $G$ is measured.   The new theory of gravity involves an
additional new dimensionless gravitational constant, and experimental data reveals this to be the
fine structure constant.  The new theory correctly predicts the globular cluster black hole masses, 
 and that  the `frame-dragging' effect is caused by
vorticity in the in-flow. The relationship of the new theory of gravity 
to General Relativity which, like Newtonian gravity, does not have the `dark matter' dynamics, 
is explained.

\newpage

\tableofcontents

\section[ Introduction ]{ Introduction
\label{section:introduction}}

The  `dark matter'  effect first came to notice some 60 years ago from observations of a mass paradox in galaxies and
galactic clusters by Oort (1932) \cite{Oort}; the dynamical estimates of the local matter density differed from that
determined from the luminosity. Zwicky (1933) \cite{Zwicky} measured the radial velocities   of galaxies in the Coma
cluster and also found that there was a mass discrepancy, with the required mass, based on the Newtonian theory of
gravity, being ten-fold greater than that deduced from the luminosities. This supposedly extra `mass' was called `dark
matter' because it produced no luminosity.  In 1959 Kahn and Woltjer \cite{KW} noticed that the relative motion of the
Andromeda galaxy and the Milky Way galaxy suggested again a ten-fold mass discrepancy.  Then Einasto \cite{Einasto},
Sizikov
\cite{Sizikov} and Freeman
\cite{Freeman}  realised that the rotation velocities in the outer regions of spiral galaxies were again much greater than
expected from the luminosity. So it began to be realised that this `dark matter' effect  was a general property of galaxies
and clusters of galaxies.  Possible `matter' interpretations for this effect have been numerous. Here we argue that it is
simply a failure of Newtonian gravity, a failure `inherited' by General Relativity. 

Newtonian gravity was based on
Kepler's Laws of motion for planets in the solar system, which were abstracted from observational data; the most famous
being that for  circular orbits the orbital speed of a planet is inversely proportional to the inverse of the square root
of the orbit radius; $v_O
\propto 1/\sqrt{ r}$. This led Newton to introduce the `universal' inverse square law of  gravity, namely that the
gravitational force between two masses is inversely proportional to the square of the separation,
\begin{equation}
F=\frac{Gm_1m_2}{r^2},
\label{eqn:Newton}\end{equation}  
which together with the acceleration equation $F=ma$, where here $a=v_O^2/r$ is the centripetal acceleration, explained
Kepler's Laws. This led to the introduction of  the gravitational acceleration vector field ${\bf g}({\bf r})$ as the
fundamental dynamical  variable for the phenomenon of gravity, and which is determined by (\ref{eqn:g1}), which relates
${\bf g}({\bf r})$ to the matter density $\rho({\bf r})$. Here $G$ is Newton's gravitational constant, the only
constant, until recent discoveries in \cite{alpha} and herein, that is involved in the phenomenon of 
gravity.  Much later Hilbert and Einstein introduced a more general theory of gravity, but which was
constrained to agree with this Newtonian theory in the appropriate limits. However  while (\ref{eqn:g1})
for ${\bf g}$  is uniquely  determined by Kepler's Laws, if we rewrite (\ref{eqn:g1}) in terms of a
velocity field
${\bf v}({\bf r},t)$ then the equation for this vector field is not uniquely determined by  Kepler's laws: a new
unique `space' self-interaction dynamical term may be incorporated that does not manifest itself in the planetary
motions of the solar system.   Numerous major developments then unfold from using this in-flow vector field 
as the fundamental degree of freedom for the phenomenon of gravity, foremost being that the new term has a strength
determined by a dimensionless constant, a second gravitational constant. Experimental data reveals
\cite{alpha} that this constant is, to within experimental error, none other than the fine structure constant
$\alpha=e^2\hbar/c \approx 1/137$. Then the most immediate result is the explanation of the so called `dark matter'
effect in spiral galaxies, though various other gravitational anomalies, as they are known,  are also now
explainable. So it turns out that the `dark matter' effect is not caused by a new so-far unidentified form of
matter, but is an effect associated with   a new feature of the phenomenon of gravity; basically gravity is a much
richer and more complex phenomenon than currently appreciated.  As discussed in more detail elsewhere \cite{NovaBook,
RC01} the velocity ${\bf v}({\bf r},t)$ field  is associated with  a restructuring and effective relative `flow' of
a quantum foam which {\it is} space; this is {\it not} a flow of something through space but is a manifestation of a
non-geometrical structure to space, with matter effectively acting as a `sink' for this quantum foam.  These deeper
insights, which are based upon an information-theoretic  modeling of reality, are discussed at length in
\cite{NovaBook, RC01}; here we mainly concentrate on various experimentally observable and observed gravitational
phenomena emerging from this new theory of gravity.  As well we show that this theory is in agreement with various
phenomena of gravity, such as precessing orbits, gravitational lensing etc, which were believed to have suggested that
General Relativity was a viable theory of gravity. We show here, of course, that the new dynamics involving the fine
structure constant is not contained in General Relativity. It is asserted here that the failure of both the Newtonian
theory and General Relativity to account for  the `dark matter' effect, and other gravitational phenomena discussed
herein, represents  a fatal flaw for both these theories; and that Newton's `universal' inverse square law
(\ref{eqn:Newton}) is not at all `universal'; it is in fact very restricted in its applicability. 

Herein  the connection of the new theory of gravity to both the Newtonian theory and to General Relativity is analysed,
but the most significant results relate to an analysis of the various  phenomena that only this new theory now
explains, including the borehole $g$ anomaly effect, the difficulties over the last 60 years in ongoing attempts
to increase the accuracy with which $G$ could be measured in Cavendish-type experiments, which are all
manifestations of the `dark matter'   effect, but which is, as explained here,  most evident in the rotation
velocity curves of spiral galaxies
\cite{alpha}. This new theory introduces a new form of quantum-foam black hole, the properties of which are
determined by the fine structure constant, and which have either `minimal' of
`non-minimal' forms. The `minimal' black holes are mandated by the in-flow into matter, and occur in all forms of
matter.  In the case of the globular clusters the effective mass of the `minimal' central black holes are computable
and found to be in agreement with recent observations.  The `non-minimal' black holes are not caused by matter and
appear to be primordial, namely residual effects of the big bang. They have  a non-inverse square law acceleration
field,  and are the cause of both the rapid formation of galaxies and of the non-Keplerian rotation dynamics of
spiral galaxies.  They have an effective `dark matter' density that falls off as almost the inverse of the square of
the distance form the black hole, as is indeed observed.  The presence of the minimal black holes  in stars affects
their internal central dynamics,  but the effect of this upon the solar neutrino flux problem has yet to be studied. 

As
already discussed elsewhere \cite{GQF,RGC,AMGE} the quantum-foam `in-flow' past the earth towards the sun has
already been shown to be present in the data from the Miller interferometer experiment of 1925/1926. That experiment
and others have also revealed the existence of gravitational waves, essentially  a flow turbulence, predicted by the
new theory of gravity, but which are very unlike those predicted, but so far unobserved, by General Relativity.

The new theory also has a `frame-dragging' effect which is being tested by the Gravity  Probe B. This
effect is caused by vorticity in the in-flow. As well the new theory has quantum-foam vortex filaments
linking, in particular, galactic black holes, and these manifest, via weak gravitational lensing, as the
recently observed `dark matter' networks.  

To avoid possible confusion it is important to understand that the special relativity effects, such as length
contractions, time dilations and mass increases, are very much a part of the new gravity theory, but that it is the
Lorentz interpretation of these effects, namely that these effects are real dynamical effects,  that is being indicated
by experiment and observation to be the correct interpretation, and not the usual non-dynamical spacetime interpretation
of these effects. In the same vein it is the failure of the Newtonian theory of gravity that is fatal for General
Relativity, and not its connection to these so-called special relativity effects.  Finally, while the `flow equations' are
classical equations, the  occurrence of $\alpha$ strongly suggests, and as predicted in \cite{NovaBook,RC01}, that this is
a manifestation of a quantum-foam substructure to space, and that we have the first experimental evidence of a quantum
theory of gravity.  As discussed here this leads to relatively easy Cavendish-type laboratory experiments that can
explore the $\alpha$-dependent aspects  of gravity - essentially laboratory quantum-gravity experiments. This
quantum-foam substructure to space also indicates an explanation of a different effect to that of `dark matter',
namely the so-called `dark energy' effect, as discussed in \cite{NovaBook}.

\section[       Gravity  as Inhomogeneous Quantum Foam In-Flow ]{ Gravity as Inhomogeneous Quantum Foam In-Flow 
\label{section:QFinflow}} 
Here we show that the Newtonian theory of gravity may be exactly re-written as a `fluid flow' system,
as can General Relativity for a class of metrics.  This `fluid' system is interpreted \cite{NovaBook,
RC01} as a classical description of a quantum foam substructure to space, and the `flow' describes the
relative motion of this quantum foam with, as we now show, gravity arising from inhomogeneities in that
flow. These inhomogeneities can be caused by an in-flow into matter, or even as inhomogeneities
produced purely by the self-interaction of space itself, as happens for instance  for the black holes.  The Newtonian
theory was originally formulated in terms of a force field, the gravitational acceleration
${\bf g}({\bf r},t)$, which is determined by  the matter density
$\rho({\bf r},t)$ according to
\begin{equation}\label{eqn:g1}
\nabla.{\bf g}=-4\pi G\rho.
\end{equation}
For $\nabla \times {\bf g}=0$ this gravitational acceleration ${\bf g}$ may be written as the
gradient of the gravitational potential $\Phi$ 
\begin{equation}{\bf g}=-{\bf \nabla}\Phi,\label{eqn:gPhi}\end{equation}  
where the  gravitational
potential\index{gravitational potential} is now determined  by 
\begin{equation}
 \nabla^2\Phi=4\pi G\rho .
\end{equation} 
 Here, as usual, $G$ is
the Newtonian  gravitational constant.  Now as $\rho\geq 0$ we can choose to have 
$\Phi
\leq 0$ everywhere if $\Phi \rightarrow 0$ at infinity. So we can introduce  ${\bf v}^2=-2\Phi \geq 0$ where 
$\bf v$ is some velocity vector field.
  Here the value of ${\bf v}^2$ is
specified, but not the direction of ${\bf v}$. Then
\begin{equation}
{\bf g}=\frac{1}{2}{\bf \nabla}({\bf v}^2)=({\bf v}.{\bf \nabla}){\bf v}+
{\bf v}\times({\bf \nabla}\times{\bf v}).
\label{eqn:f1}
\end{equation} 
For zero-vorticity (irrotational) flow 
${\bf \omega}={\bf \nabla} \times {\bf v}={\bf 0}$. Then ${\bf g}$ is
the usual Euler expression for the  acceleration of a fluid element in a
time-independent or stationary fluid flow.   If the flow is time dependent that expression
is expected to become
 \begin{equation}{\bf g}=\displaystyle{\frac{\partial {\bf v}}{\partial t}}+({\bf v}.{\bf \nabla}){\bf
v}=\displaystyle{\frac{d {\bf v}}{d t}},
\label{eqn:f2}\end{equation}
which has given rise to the total derivative of ${\bf v}$ familiar from fluid mechanics. This equation  is then to
be accompanied by the `Newtonian equation' for the flow field
\begin{equation}
\frac{1}{2}\nabla^2({\bf v}^2)=-4\pi G\rho,
\label{eqn:f3a}\end{equation}
but to be consistent with (\ref{eqn:f2}) in the case of a time-dependent matter density this equation
should be generalised to 
\begin{equation}
\frac{\partial }{\partial t}(\nabla.{\bf v})+\nabla.(({\bf
v}.{\bf \nabla}){\bf v})=-4\pi G\rho.
\label{eqn:f3}\end{equation}
This exhibits the fluid flow form of  Newtonian gravity in the case of zero vorticity $\nabla \times
{\bf v}=0$.  For zero vorticity (\ref{eqn:f3}) determines both the magnitude and direction of the
velocity field, for in this case we can write ${\bf v}=\nabla u$, where $u({\bf r},t)$ is a scalar
velocity potential, and in terms of $u({\bf r},t)$ (\ref{eqn:f3})  specifies uniquely the time evolution
of $u({\bf r},t)$.  Note that (\ref{eqn:f2})  and (\ref{eqn:f3})  are exactly equivalent to
(\ref{eqn:g1}) for the acceleration field ${\bf g}$, and so   within the fluid flow formalism
(\ref{eqn:f2}) and (\ref{eqn:f3}) are together equivalent to the Universal Inverse Square Law for
 ${\bf g}$, and so both are equally valid as regards the numerous experimental and observational checks
of the acceleration field ${\bf g}$  formalism, particularly the Keplerian rotation velocity law. So we
appear to have two equivalent formalisms for the same phenomenon. Indeed for a stationary spherically symmetric
distribution of matter of total mass
$M$ the velocity field outside of the matter 
\begin{equation}
{\bf v}({\bf r})=-\sqrt{\frac{2GM}{r}}\hat{\bf r},
\label{eqn:vfield}\end{equation}
satisfies (\ref{eqn:f3}) and reproduces the inverse square law form for ${\bf g}$ using (\ref{eqn:f2}): 
\begin{equation}
{\bf g}=-\frac{GM}{r^2}\hat{\bf r}.
\label{eqn:InverseSqLaw}\end{equation} 

So the immediate questions that arise are (i)  can the two formalisms be distinguished experimentally, and (ii)
can the velocity field formalism be generalised, leading  to new gravitational phenomena?   To answer these questions we
note that 
\begin{enumerate}
\item The velocity flow field of some 430km/s in the direction (Right Ascension $=5.2^{hr}$, Declination$ =
-67^0$)  has been detected in several experiments, as described in considerable detail in 
\cite{NovaBook,RGC,AMGE}. The major component of that flow is related to a galactic flow, presumably  within the Milky
Way and  the local galactic cluster, but a smaller component of some 50km/s being the  flow past the earth
towards the sun has also recently been revealed in the data.

\item In terms of the velocity field formalism (\ref{eqn:f3}) a unique term may be added that does not affect
observations  within the solar system, such as encoded in Kepler's laws, but outside of that special case the new term
causes effects which vary from small to extremely large.  This term will be shown herein to cause those effects that have
been mistakenly called the `dark matter' effect. 

\item Eqn.(\ref{eqn:f3}) and its generalisations have time-dependent solutions even when the matter density is not
time-dependent. These are a form of flow turbulence, a gravitational wave effect, and they have also been detected, as
discussed in \cite{NovaBook,RGC,AMGE}. 

\item  The need for a further generalisation of the flow equations will be argued for, and this in particular
includes flow vorticity that leads to a non-spacetime explanation of the `frame-dragging' effect, and of the `dark
matter' network observed using the weak gravitational lensing technique.
\end{enumerate}

First let us consider the arguments that lead to a generalisation of (\ref{eqn:f3}). The simplest generalisation
is
\begin{equation}
\frac{\partial }{\partial t}(\nabla.{\bf v})+\nabla.(({\bf
v}.{\bf \nabla}){\bf v})+C({\bf v})=-4\pi G\rho,
\label{eqn:f3extend}\end{equation}
where
\begin{equation}
C({\bf v})=\displaystyle{\frac{\alpha}{8}}((tr D)^2-tr(D^2)),
\label{eqn:Cdefn1}\end{equation} and
\begin{equation}
D_{ij}=\frac{1}{2}(\frac{\partial v_i}{\partial x_j}+\frac{\partial v_j}{\partial x_i})
\label{eqn:Ddefn1}\end{equation}
is the symmetric part of the rate of strain tensor $\partial v_i/\partial x_j$, and $\alpha$ is a dimensionless
constant - a new gravitational constant in addition to $G$.  It is possible to check that for the in-flow in
(\ref{eqn:vfield}) $C({\bf v})=0$. This is a feature that uniquely determines the form of $C({\bf v})$.  This means
that effects  caused by this new term are not manifest in the planetary motions that formed the basis of Kepler's
phenomenological laws and that then lead to Newton's theory of gravity.  As we shall see the value of $\alpha$
determined from experimental data is found to be the fine structure constant, to within experimental error. As 
well, as discussed in Sect.\ref{section:dmas} and extensively after that,
(\ref{eqn:f3extend}) predicts precisely the  so-called `dark matter' effect, with the effective `dark matter'
density defined by  
\begin{equation}
\rho_{DM}({\bf r})=\frac{\alpha}{32\pi G}( (tr D)^2-tr(D^2)).  
\label{eqn:DMdensity0}\end{equation} 
So the explanation of the `dark matter'
effect becomes apparent  once we use the velocity field  formulation of gravity. However  (\ref{eqn:f3extend}) must
be further generalised to include  (i)  the velocity of absolute motion  of the matter
components with respect to the local quantum foam system, and (ii) vorticity effects.

For these further generalisations we need to be  precise by what is meant by the velocity field
 ${\bf v}({\bf r},t)$. To be specific and also to define a possible measurement procedure
 we can choose to  use the Cosmic Microwave Background (CMB)\index{Cosmic Microwave
Background (CMB)} frame of reference for that purpose, as this is itself easy to establish. However that does not
imply that the  CMB frame is the local `quantum-foam' rest frame. Relative to the CMB frame and using the local
absolute motion detection techniques described in \cite{NovaBook,RGC,AMGE}, or more modern techniques that are under
development, ${\bf v}({\bf r},t)$ may be measured in the neighbourhood of the observer.   Then an `object' at
location ${\bf r}_0(t)$ in the CMB frame has velocity  
 ${\bf v}_0(t)=d{\bf r}_0(t)/dt$ 
with respect to that  frame.  We then define 
\begin{equation}
{\bf v}_R({\bf r}_0(t),t) ={\bf v}_0(t) - {\bf v}({\bf r}_0(t),t),
\label{eqn:18}
\end{equation}
as the velocity of the object relative to the quantum foam at the location of the object.  However this absolute
velocity of matter ${\bf v}_R(t)$ does not appear in (\ref{eqn:f3extend}), and so not only is that equation lacking
vorticity effects, it presumably is only an approximation for when the matter has a negligible speed of absolute
motion with respect to the local quantum foam.  To introduce  the vector ${\bf v}_R(t)$ we need to construct a
2nd-rank tensor generalisation of   (\ref{eqn:f3extend}), and the simplest form is 
\begin{eqnarray}
&&\frac{d D_{ij}}{dt}+\frac{\delta_{ij}}{3}tr(D^2) +\frac{tr D}{2}
(D_{ij}-\frac{\delta_{ij}}{3}tr D)\nonumber \\ &&+\frac{\delta_{ij}}{3}\frac{\alpha}{8}((tr D)^2 -tr(D^2))=-4\pi
G\rho(\frac{\delta_{ij}}{3}+\frac{v^i_{R}v^j_{R}}{2c^2}+..)  ,  \mbox{ } i,j=1,2,3.
\label{eqn:f3general}\end{eqnarray}
which uses the total derivative of the $D_{ij}$ tensor in (\ref{eqn:Ddefn1}). Because of its tensor structure we
can  now include the direction of absolute motion of the matter density with respect to the quantum foam, with the
scale of that given by
$c$, which is the speed of light relative to the quantum foam. The superscript notation for the components of   ${\bf
v}_R(t)$ is for convenience only, and has no other significance.  The trace of (\ref{eqn:f3general}), using the
identity
\begin{equation}
({\bf v}.\nabla)(tr D)=\frac{1}{2}\nabla^2({\bf v}^2)
-tr(D^2)-\frac{1}{2}(\nabla\times{\bf v})^2+ {\bf v}.\nabla\times(\nabla\times{\bf v}),
\label{eqn:identity}\end{equation}
gives, for zero vorticity,
\begin{equation}
\frac{\partial }{\partial t}(\nabla.{\bf v})+\nabla.(({\bf
v}.{\bf \nabla}){\bf v})+C({\bf v})=-4\pi G\rho(1+\frac{v_R^2}{2c^2}+..),
\label{eqn:f3p}\end{equation}
which is (\ref{eqn:f3extend}) in the limit $v_R\rightarrow 0$.  As well the off-diagonal terms,  $i\neq j$,
are satisfied, to   $O(v_R^iv_R^j/c^2)$, for the in-flow velocity field in (\ref{eqn:vfield}). The
conjectured form of the RHS of (\ref{eqn:f3p}) is, to 
$O(v_R^2/c^2)$,  based on the Lorentz contraction effect for the matter density, with
$\rho$ defined as the matter density  if the matter were at rest with respect to the quantum foam. Hence, because of
(\ref{eqn:f3p}),  (\ref{eqn:f3general})  is in agreement with Keplerian orbits for the solar system with the
velocity field given by   (\ref{eqn:vfield}).

We now consider a further generalisation of (\ref{eqn:f3general}) to include vorticity effects, namely 
\begin{eqnarray}
&&\frac{d D_{ij}}{dt}+ \frac{\delta_{ij}}{3}tr(D^2) + \frac{tr D}{2}
(D_{ij}-\frac{\delta_{ij}}{3}tr D)\nonumber \\ &&+\frac{\delta_{ij}}{3}\frac{\alpha}{8}((tr
D)^2 -tr(D^2)) -(D\Omega-\Omega D)_{ij}\nonumber \\&&\mbox{\ \ \ \ \ \ \ \ \  }=-4\pi
G\rho(\frac{\delta_{ij}}{3}+\frac{v^i_{R}v^j_{R}}{2c^2}+..), 
\mbox{ \ \ } i,j=1,2,3, 
\label{eqn:f3vorticitya}\end{eqnarray}
\begin{equation}\nabla \times(\nabla\times {\bf v}) =\frac{8\pi G\rho}{c^2}{\bf v}_R,
\label{eqn:f3vorticityb}\end{equation}
 where
\begin{equation}
\Omega_{ij}=\frac{1}{2}(\frac{\partial v_i}{\partial x_j}-\frac{\partial v_j}{\partial
x_i})=-\frac{1}{2}\epsilon_{ijk}\omega_k=-\frac{1}{2}\epsilon_{ijk}(\nabla\times {\bf v})_k
\label{eqn:Omegadefn}\end{equation}
is the antisymmetric part of the rate of strain tensor $\partial v_i/\partial x_j$, which is the vorticity vector
field $\omega$ in tensor form. The term
$(D\Omega-\Omega D)_{ij}$ allows the vorticity  vector field to be coupled to the symmetric tensor $D_{ij}$
dynamics. Again the vorticity is generated by absolute motion of the matter density with respect to the local
quantum foam. Eqns (\ref{eqn:f3vorticitya}) and (\ref{eqn:f3vorticityb}) now permit the time evolution of the
velocity field  to be determined. Note that the vorticity equation in (\ref{eqn:f3vorticityb}) may be explicitly
solved, for it may be written as
\begin{equation}
\nabla(\nabla.{\bf v})-\nabla^2 {\bf v}=\frac{8\pi G\rho}{c^2}{\bf v}_R,
\label{veqn}\end{equation}
which gives, using
\begin{equation}
\nabla^2\left(\frac{1}{|{\bf r} - {\bf r^\prime}|}  \right)=-4\pi\delta({\bf r} - {\bf r^\prime}),
\label{eqn:deltafnidentity}\end{equation}
\begin{equation}
{\bf v}({\bf r},t)=\frac{2G}{c^2}\int d^3 r^\prime \frac{\rho({\bf r}^\prime,t)}{|{\bf r}-{\bf r}^\prime|}{\bf
v}_R({\bf r}^\prime,t)-
\frac{1}{4\pi}\int d^3 r^\prime \frac{1}{|{\bf r}-{\bf r}^\prime|}\nabla(\nabla.{\bf v}({\bf
r}^\prime,t)).
\label{eqn:BS1}\end{equation}
This suggests that ${\bf v}({\bf r},t)$ is now determined solely by the vorticity equation. However  (\ref{eqn:BS1})
is misleading, as  (\ref{eqn:f3vorticityb}) only specifies the vorticity, and taking the $\nabla \times$ of  
(\ref{eqn:BS1}) we obtain
\begin{equation}
\omega({\bf r},t)=\nabla\times{\bf v}({\bf r},t)
=\frac{2G}{c^2}\int d^3 r^\prime \frac{\rho({\bf r}^\prime,t)}
{|{\bf r}-{\bf r}^\prime|^3}{\bf v}_R({\bf r}^\prime,t)\times({\bf r}-{\bf r}^\prime)+\nabla\psi,
\label{eqn:BS2}\end{equation} 
which is the Biot-Savart form for the vorticity, with the additional term being the homogeneous solution.  Then
(\ref{eqn:f3vorticitya}) becomes an integro-differential equation for the velocity field, with $\psi$ determined by
self-consistency.  As we shall see in
Sect.\ref{section:lense} (\ref{eqn:BS2}) explains the so-called  `frame-dragging'  effect in terms of
this vorticity in the in-flow.  Of course (\ref{eqn:f3vorticitya}) and (\ref{eqn:BS2}) only make sense if  
${\bf v}_R({\bf r},t)$ for the matter at location ${\bf r}$  is specified. We now consider the special case where
the matter is subject only to the effects of motion with respect to the quantum-foam velocity-field
inhomogeneities and  variations in time, which causes a `gravitational' acceleration.

We also not that (\ref{eqn:f3vorticitya}) and (\ref{eqn:f3vorticityb}) need to be further generalised  to take
account of the cosmological-scale effects, namely that the spatial system is compact and growing, as discussed in
\cite{NovaBook}.

\section{ Geodesics \label{section:geodesics}}

Process Physics \cite{NovaBook} leads to the Lorentzian interpretation\index{Lorentzian interpretation} of
so called `relativistic effects'.  This means that the speed of light \index{speed of light} is only `c' with
respect to the quantum-foam system, and that time dilation effects for clocks and length contraction effects for
rods are caused by the motion of clocks and rods relative to the quantum foam. So these effects are real dynamical
effects caused by motion through the quantum foam, and are not to be interpreted as non-dynamical spacetime
effects as suggested by Einstein.  To arrive at the dynamical description of the various effects of the quantum foam
we shall introduce conjectures that essentially lead to a phenomenological description of these effects. In
the future we expect to be able to derive this dynamics directly from the Quantum Homotopic Field Theory (QHFT)
 that describes the quantum foam system \cite{NovaBook}. Here we shall conjecture that the path of an object
through an inhomogeneous and time-varying quantum-foam is determined, at a classical level,  by a variational
principle, namely that  the travel time is extremised for the physical path ${\bf r}_0(t)$. The travel time is
defined by 
\begin{equation}
\tau[{\bf r}_0]=\int dt \left(1-\frac{{\bf v}_R^2}{c^2}\right)^{1/2},
\label{eqn:f4}
\end{equation}  
with ${\bf v}_R$ given by (\ref{eqn:18}). So the trajectory will be independent of the mass of the object,
corresponding to the equivalence principle.  Under a deformation of the trajectory  ${\bf r}_0(t) \rightarrow  {\bf
r}_0(t) +\delta{\bf r}_0(t)$,
${\bf v}_0(t) \rightarrow  {\bf v}_0(t) +\displaystyle\frac{d\delta{\bf r}_0(t)}{dt}$,  and we also
have
\begin{equation}\label{eqn:G2}
{\bf v}({\bf r}_0(t)+\delta{\bf r}_0(t),t) ={\bf v}({\bf r}_0(t),t)+(\delta{\bf
r}_0(t).{\bf \nabla}) {\bf v}({\bf r}_0(t))+... 
\end{equation}
Then
\begin{eqnarray}\label{eqn:G3}
\delta\tau&=&\tau[{\bf r}_0+\delta{\bf r}_0]-\tau[{\bf r}_0]  \nonumber\\
&=&-\int dt \:\frac{1}{c^2}{\bf v}_R. \delta{\bf v}_R\left(1-\displaystyle{\frac{{\bf
v}_R^2}{c^2}}\right)^{-1/2}+...\nonumber\\
&=&\int dt\frac{1}{c^2}\left({\bf
v}_R.(\delta{\bf r}_0.{\bf \nabla}){\bf v}-{\bf v}_R.\frac{d(\delta{\bf
r}_0)}{dt}\right)\left(1-\displaystyle{\frac{{\bf v}_R^2}{c^2}}\right)^{-1/2}+...\nonumber\\ 
&=&\int dt \frac{1}{c^2}\left(\frac{{\bf v}_R.(\delta{\bf r}_0.{\bf \nabla}){\bf v}}{ 
\sqrt{1-\displaystyle{\frac{{\bf
v}_R^2}{c^2}}}}  +\delta{\bf r}_0.\frac{d}{dt} 
\frac{{\bf v}_R}{\sqrt{1-\displaystyle{\frac{{\bf
v}_R^2}{c^2}}}}\right)+...\nonumber\\
&=&\int dt\: \frac{1}{c^2}\delta{\bf r}_0\:.\left(\frac{({\bf v}_R.{\bf \nabla}){\bf v}+{\bf v}_R\times({\bf
\nabla}\times{\bf v})}{ 
\sqrt{1-\displaystyle{\frac{{\bf
v}_R^2}{c^2}}}}  +\frac{d}{dt} 
\frac{{\bf v}_R}{\sqrt{1-\displaystyle{\frac{{\bf
v}_R^2}{c^2}}}}\right)+...
\end{eqnarray}
  Hence a 
trajectory ${\bf r}_0(t)$ determined by $\delta \tau=0$ to $O(\delta{\bf r}_0(t)^2)$ satisfies 
\begin{equation}\label{eqn:G4}
\frac{d}{dt} 
\frac{{\bf v}_R}{\sqrt{1-\displaystyle{\frac{{\bf v}_R^2}{c^2}}}}=-\frac{({\bf
v}_R.{\bf \nabla}){\bf v}+{\bf v}_R\times({\bf
\nabla}\times{\bf v})}{ 
\sqrt{1-\displaystyle{\frac{{\bf v}_R^2}{c^2}}}}.
\label{eqn:vReqn}\end{equation}
Let us now write this in a more explicit form.  This will
also allow the low speed limit to be identified.   Substituting ${\bf
v}_R(t)={\bf v}_0(t)-{\bf v}({\bf r}_0(t),t)$ and using 
\begin{equation}\label{eqn:G5}
\frac{d{\bf v}({\bf r}_0(t),t)}{dt}=\frac{\partial {\bf v}}{\partial t}+({\bf v}_0.{\bf \nabla}){\bf
v},
\end{equation}
we obtain
\begin{equation}\label{eqn:G6}
\frac{d}{dt} 
\frac{{\bf v}_0}{\sqrt{1-\displaystyle{\frac{{\bf v}_R^2}{c^2}}}}={\bf v}
\frac{d}{dt}\frac{1}{\sqrt{1-\displaystyle{\frac{{\bf v}_R^2}{c^2}}}}+\frac{\displaystyle{\frac{\partial {\bf
v}}{\partial t}}+({\bf v}.{\bf \nabla}){\bf v}+({\bf \nabla}\times{\bf v})\times{\bf v}_R}{ 
\displaystyle{\sqrt{1-\frac{{\bf v}_R^2}{c^2}}}}.
\end{equation}
Then in the low speed limit  $v_R \ll c $   we  obtain
\begin{equation}{\label{eqn:G7}}
\frac{d{\bf v}_0}{dt}=\frac{\partial {\bf v}}{\partial t}+({\bf v}.{\bf
\nabla}){\bf v}+({\bf \nabla}\times{\bf v})\times{\bf v}_R,
\end{equation}
which agrees with the  fluid flow form suggested in  (\ref{eqn:f2}) for zero vorticity (${\bf \nabla}\times{\bf
v}=0$), but introduces a new vorticity effect for the gravitational acceleration.  The last term in (\ref{eqn:G7}) is
relevant to the `frame-dragging' effect and  to the Allais eclipse effect. Hence (\ref{eqn:G6}) is a
generalisation of (\ref{eqn:f2}) to include  Lorentzian dynamical effects, for  in (\ref{eqn:G6})  we can
multiply both sides by the rest mass  $m_0$ of the object, and   then (\ref{eqn:G6}) involves 
\begin{equation}
m({\bf v}_R) =\frac{m_0}{\sqrt{1-\displaystyle{\frac{{\bf v}_R^2}{c^2}}}},
\label{eqn:G8}\end{equation}
the so called `relativistic' mass\index{relativistic mass}, and (\ref{eqn:G6}) acquires the form
\begin{equation}\frac{d}{dt}(m({\bf v}_R){\bf v}_0)={\bf F},\end{equation} where
${\bf F}$ is an effective `force' caused by the inhomogeneities and time-variation of the flow.  This is
essentially Newton's 2nd Law of Motion in the case of gravity only. That $m_0$ cancels is the equivalence principle, 
and which acquires a simple explanation in terms of the flow.  Note that the
occurrence of
$1/\sqrt{1-\frac{{\bf v}_R^2}{c^2}}$ will lead to the precession of the perihelion of elliptical planetary orbits,
and also to horizon effects wherever  $|{\bf v}| = c$: the region where  $|{\bf v}| < c$ is
inaccessible from the region where $|{\bf v}|>c$.  Also (\ref{eqn:f4}) is easily used to determine the 
clock rate offsets in the GPS satellites\index{Global Positioning System (GPS)}, when the in-flow is given by
(\ref{eqn:vfield}). So the fluid flow dynamics in  (\ref{eqn:f3vorticitya}) and (\ref{eqn:BS2}) and the gravitational
dynamics for the matter in  (\ref{eqn:vReqn}) now form a closed system.  
This system of equations is a considerable
generalisation from that of Newtonian gravity, and would appear to be very different from the curved spacetime
formalism of General Relativity. However we now show that General Relativity leads to a very similar system of
equations, but with one important exception, namely that the `dark matter' `quantum-foam' dynamics is missing from the
Hilbert-Einstein theory of gravity. 

The above may  be modified when the `object' is a massless photon, and the
corresponding result leads to the gravitational lensing effect. But not only will ordinary matter produce such
lensing, but the effective `dark matter' density will also do so, and that is relevant to the recent observation by
the weak lensing technique of the so-called `dark matter' networks, in Sect.\ref{section:vortex}.

\section[  General Relativity and the In-Flow Process ]{  General Relativity and the In-Flow Process 
\label{section:general}} 

Eqn.(\ref{eqn:f4})  involves various absolute quantities such  as the absolute velocity of an object
relative to the  quantum foam and the absolute speed
$c$ also relative to the foam, and of course absolute velocities are excluded from the General Relativity (GR)
formalism.  However (\ref{eqn:f4}) gives (with $t=x_0^0$)
\begin{equation}
d\tau^2=dt^2-\frac{1}{c^2}(d{\bf r}_0(t)-{\bf v}({\bf r}_0(t),t)dt)^2=
g_{\mu\nu}(x_0 )dx^\mu_0 dx^\nu_0,
\label{eqn:24}\end{equation}
which is  the   Panlev\'{e}-Gullstrand form of the metric $g_{\mu\nu}$
\index{metric $g_{\mu\nu}$} 
\cite{PP, AG} for GR.  We emphasize that the
absolute velocity
${\bf v}_R$  has been measured, and so the foundations of GR as usually stated are invalid. 
Here we look closely at the GR formalism when the metric has the form in (\ref{eqn:24}), appropriate to a 
velocity field  formulation of gravity.  In GR the  metric tensor $g_{\mu\nu}(x)$, specifying the geometry
of the spacetime construct, is determined by
\begin{equation}
G_{\mu\nu}\equiv R_{\mu\nu}-\frac{1}{2}Rg_{\mu\nu}=\frac{8\pi G}{c^2} T_{\mu\nu},
\label{eqn:32}\end{equation}
where  $G_{\mu\nu}$ is  the Einstein tensor, $T_{\mu\nu}$ is the  energy-momentum tensor,
$R_{\mu\nu}=R^\alpha_{\mu\alpha\nu}$ and
$R=g^{\mu\nu}R_{\mu\nu}$ and
$g^{\mu\nu}$ is the matrix inverse of $g_{\mu\nu}$. The curvature tensor is
\begin{equation}
R^\rho_{\mu\sigma\nu}=\Gamma^\rho_{\mu\nu,\sigma}-\Gamma^\rho_{\mu\sigma,\nu}+
\Gamma^\rho_{\alpha\sigma}\Gamma^\alpha_{\mu\nu}-\Gamma^\rho_{\alpha\nu}\Gamma^\alpha_{\mu\sigma},
\label{eqn:curvature}\end{equation}
where $\Gamma^\alpha_{\mu\sigma}$ is the affine connection
\begin{equation}
\Gamma^\alpha_{\mu\sigma}=\frac{1}{2} g^{\alpha\nu}\left(\frac{\partial g_{\nu\mu}}{\partial x^\sigma}+
\frac{\partial g_{\nu\sigma}}{\partial x^\mu}-\frac{\partial g_{\mu\sigma}}{\partial x^\nu} \right).
\label{eqn:affine}\end{equation}
In this formalism the trajectories of test objects are determined by
\begin{equation}
\Gamma^\lambda_{\mu\nu}\frac{dx^\mu}{d\tau}\frac{dx^\nu}{d\tau}+\frac{d^2x^\lambda}{d\tau^2}=0,
\label{eqn:33}\end{equation}
 which is equivalent to extremising the functional
\begin{equation}
\tau[x]=\int dt\sqrt{g^{\mu\nu}\frac{dx^{\mu}}{dt}\frac{dx^{\nu}}{dt}},
\label{eqn:path}\end{equation}
with respect  to the path $x[t]$. This is precisely equivalent to (\ref{eqn:f4}).  

In the case   of a spherically symmetric mass $M$ the well known   solution of
(\ref{eqn:32}) outside of that mass    is the external-Schwarzschild metric
\begin{equation}
d\tau^2=(1-\frac{2GM}{c^2r})dt^{ 2}-
\frac{1}{c^2}r^{ 2}(d\theta^2+\sin^2(\theta)d\phi^2)-\frac{dr^{ 2}}{c^2(1-\frac{\displaystyle
2GM}{\displaystyle c^2r})}.
\label{eqn:SM}\end{equation}
This solution is the basis of various experimental checks of General Relativity in which the spherically symmetric
mass is either the sun or the earth.  The four tests are: the gravitational redshift, the bending of light, the
precession of the perihelion of Mercury, and the time delay of radar signals.

 However the solution (\ref{eqn:SM}) is in fact
completely equivalent to the in-flow interpretation of Newtonian gravity.  Making the change of variables
$t\rightarrow t^\prime$ and
$\bf{r}\rightarrow {\bf r}^\prime= {\bf r}$ with
\begin{equation}
t^\prime=t+
\frac{2}{c}\sqrt{\frac{2GMr}{c^2}}-\frac{4GM}{c^2}\mbox{tanh}^{-1}\sqrt{\frac{2GM}{c^2r}},
\label{eqn:37}\end{equation}
the Schwarzschild solution (\ref{eqn:SM}) takes the form
\begin{equation}
d\tau^2=dt^{\prime 2}-\frac{1}{c^2}(dr^\prime+\sqrt{\frac{2GM}{r^\prime}}dt^\prime)^2-\frac{1}{c^2}r^{\prime
2}(d\theta^{\prime 2}+\sin^2(\theta^\prime)d\phi^{\prime 2}),
\label{eqn:PG}\end{equation}
which is exactly  the  Panlev\'{e}-Gullstrand form of the metric $g_{\mu\nu}$ 
\cite{PP, AG} in (\ref{eqn:24})
 with the velocity field given exactly  by the Newtonian form in (\ref{eqn:vfield}).   In which case the geodesic
equation (\ref{eqn:33}) of test objects in the Schwarzschild metric is equivalent to solving (\ref{eqn:G6}).  
This choice of coordinates corresponds to a particular frame of reference in which the test object has velocity
${\bf v}_R={\bf v}-{\bf v}_0$ relative to the in-flow field ${\bf v}$, as seen in (\ref{eqn:f4}).    This results
shows that the Schwarzschild metric in GR is completely equivalent to Newton's inverse square law: GR in
this case is nothing more than Newtonian gravity in disguise.  So the so-called `tests' of GR were nothing more than
a test of the geodesic equation, where most simply this is seen to determine the motion of an object relative to an
absolute local frame of reference - the quantum foam frame. 

 It is conventional wisdom for practitioners in  General
Relativity  to regard the choice of coordinates or frame of reference to be entirely arbitrary and having no physical
significance:  no observations should be possible that can detect and measure ${\bf v}_R$.  This `wisdom' is based
on two  beliefs (i) that all attempts to detect ${\bf v}_R$, namely the detection of absolute motion, have
failed, and that (ii)  the existence of absolute motion is incompatible with the many successes of both the
Special Theory of Relativity and of the General Theory of Relativity.  Both of these beliefs are demonstrably
false, see \cite{NovaBook,GQF}. 

The results in this section suggest, just as for Newtonian
gravity, that the Einstein General Relativity is nothing more than the dynamical equations for a velocity flow field 
${\bf v}({\bf r },t)$.  Hence  the non-flat spacetime\index{spacetime} construct appears to be merely an
unnecessary  artifact of the Einstein measurement protocol, which in turn was motivated by the mis-reporting of the
results of the Michelson-Morley experiment \cite{NovaBook,AMGE}. The putative successes of General Relativity should
thus be considered as an insight into  the fluid flow dynamics of the quantum foam system, rather than any
confirmation of the validity of the spacetime formalism, and it was this insight that in \cite{NovaBook,GQF} led, in
part, to the flow dynamics in (\ref{eqn:f3vorticitya}) and (\ref{eqn:f3vorticityb}).   
Let us therefore  substitute the metric 
\begin{equation}
d\tau^2=g_{\mu\nu}dx^\mu dx^\nu=dt^2-\frac{1}{c^2}(d{\bf r}(t)-{\bf v}({\bf r}(t),t)dt)^2,
\label{eqn:PGmetric}\end{equation}
into (\ref{eqn:32})  using  (\ref{eqn:affine})  and (\ref{eqn:curvature}). This metric involves the arbitrary
time-dependent velocity field  ${\bf v}({\bf r},t)$.  The various components of the
Einstein tensor are then found to be
\begin{eqnarray}\label{eqn:G}
G_{00}&=&\sum_{i,j=1,2,3}v_i\mathcal{G}_{ij}
v_j-c^2\sum_{j=1,2,3}\mathcal{G}_{0j}v_j-c^2\sum_{i=1,2,3}v_i\mathcal{G}_{i0}+c^2\mathcal{G}_{00}, 
\nonumber\\ G_{i0}&=&-\sum_{j=1,2,3}\mathcal{G}_{ij}v_j+c^2\mathcal{G}_{i0},   \mbox{ \ \ \ \ } i=1,2,3.
\nonumber\\ G_{ij}&=&\mathcal{G}_{ij},   \mbox{ \ \ \ \ } i,j=1,2,3.
\end{eqnarray}
where the  $\mathcal{G}_{\mu\nu}$ are  given by
\begin{eqnarray}\label{eqn:GT}
\mathcal{G}_{00}&=&\frac{1}{2}((trD)^2-tr(D^2)), \nonumber\\
\mathcal{G}_{i0}&=&\mathcal{G}_{0i}=-\frac{1}{2}(\nabla\times(\nabla\times{\bf v}))_i,   \mbox{ \ \ \ \ }
i=1,2,3.\nonumber\\ 
\mathcal{G}_{ij}&=&
\frac{d}{dt}(D_{ij}-\delta_{ij}trD)+(D_{ij}-\frac{1}{2}\delta_{ij}trD)trD\nonumber\\ & &
-\frac{1}{2}\delta_{ij}tr(D^2)-(D\Omega-\Omega D)_{ij},  \mbox{ \ \ \ \ } i,j=1,2,3.
\end{eqnarray}
In vacuum, with $T_{\mu\nu}=0$, we find from (\ref{eqn:32}) and (\ref{eqn:G}) that $G_{\mu\nu}=0$ implies that  
$\mathcal{G}_{\mu\nu}=0$. This system of equations is thus very similar to the in-flow dynamics in
(\ref{eqn:f3vorticitya}) and (\ref{eqn:f3vorticityb}), except that in vacuum GR, for the  Panlev\'{e}-Gullstrand
metric, demands that
\begin{equation}
((trD)^2-tr(D^2))=0.
\label{eqg:DMGR}\end{equation}  
This simply corresponds to the fact that GR does not permit the `dark matter' effect, namely that
$\rho_{DM}=0,$ according to (\ref{eqn:DMdensity0}), and this happens because GR was forced to agree with Newtonian
gravity, in the appropriate limits, and that theory also has no such effect. As well in GR the energy-momentum
tensor
$T_{\mu\nu}$ is not permitted to make any reference to absolute linear motion of the matter; only  the relative
motion of matter or absolute rotational motion is permitted.

It is very significant to note that the above exposition of the GR formalism for  the  Panlev\'{e}-Gullstrand
metric is exact. Then taking the trace of the $\mathcal{G}_{ij}$ equation in (\ref{eqn:GT}) we obtain,
also exactly, and again using the identity in  (\ref{eqn:identity}), and in the case of zero vorticity,
and outside of matter so that
$T_{\mu\nu}=0$,
\begin{equation}
\frac{\partial }{\partial t}(\nabla.{\bf v})+\nabla.(({\bf
v}.{\bf \nabla}){\bf v})=0,
\label{eqn:f3vacuum}\end{equation}
which is the Newtonian `velocity field' formulation of Newtonian gravity outside of matter.  
This should have been expected as it corresponds to  the previous observation that  `Newtonian in-flow' velocity
field is exactly equivalent to the external Schwarzschild metric. So again we see  the extreme paucity of new
physics in the GR formalism:  all the key tests of GR are now seen to amount to a test {\it only} of $\delta
\tau[x]/\delta x^\mu = 0$,  when the in-flow field is given by  (\ref{eqn:GT}), and which
is nothing more than Newtonian gravity. Of course Newtonian gravity was itself merely based upon observations within
the solar system, and this was too special to have revealed key aspects of gravity. Hence, despite popular
opinion, the GR formalism is  based upon  very poor evidence. Indeed there is only one 
definitive confirmation of the GR formalism apart from  the misleading external-Schwarzschild metric cases, namely
the observed decay of  the binary pulsar orbital motion, for only in this case is the metric non-Schwarzschild, and
therefore non-Newtonian.  However the new theory of
gravity also leads to the decay of orbits, and on the grounds of dimensional analysis we would expect
comparable predictions.  So  GR is not unique in predicting orbital decay.

\section[ The  `Dark Matter' Effect]{ The  `Dark Matter'  Effect\label{section:dmas}}

We now make more explicit the `dark matter' effect in a form that will be extensively analysed in the following
sections. Restricting  the flow dynamics to that of a matter system approximately at rest with respect to the
quantum foam system, and also neglecting vorticity effects, (\ref{eqn:f3vorticitya}) and (\ref{eqn:f3vorticityb})
simplify to  (\ref{eqn:f3extend}), with the key   
$C({\bf v})$ term defined in (\ref{eqn:Cdefn1}).    In this case we have 
\begin{equation}\label{eqn:ga}
{\bf g}=\frac{\partial {\bf v}}{\partial
t}+( {\bf v}.\nabla){\bf v},\end{equation} 
and then (\ref{eqn:f3extend}) gives
\begin{equation}\label{eqn:g2}
\nabla.{\bf g}=-4\pi G\rho-C({\bf v})=-4\pi G\rho-4\pi G \rho_{DM},
\end{equation}
after writing  the new term as $C({\bf v})=4\pi G \rho_{DM}$, with
\begin{equation}
\rho_{DM}({\bf r})=\frac{\alpha}{32\pi G}( (tr D)^2-tr(D^2)).  
\label{eqn:DMdensity1}\end{equation} 
So we see that 
$\rho_{DM}$ would act as an effective matter density, and it is demonstrated later that it is the consequences of
this term which have been misinterpreted as `dark matter'. Note however $\rho_{DM}$ is not positive definite. We see
that this effect is actually the consequence of quantum foam effects within the new proposed dynamics for gravity,
and which becomes apparent particularly in spiral galaxies.  
With $\nabla\times{\bf v}=0$ we can write ${\bf v}=\nabla u$, and (\ref{eqn:f3extend}) has the form
\begin{equation}
\nabla^2\left(\frac{\partial u}{\partial t}+\frac{1}{2}(\nabla u)^2\right)=-4\pi G\rho-C(\nabla u({\bf r})).
\label{eqn:ueqn}\end{equation}
Then noting (\ref{eqn:deltafnidentity})
 we see that (\ref{eqn:ueqn}) has the non-linear integro-differential equation form
\begin{equation}\label{eqn:ueqn2}
\frac{\partial u({\bf r},t)}{\partial t}=-\frac{1}{2}(\nabla u({\bf r},t))^2+\frac{1}{4\pi}\int d^3
r^\prime\frac{C(\nabla u({\bf r}^\prime,t))}{|{\bf r}-{\bf r}^\prime|}-\Phi({\bf r},t),
\end{equation}
where $\Phi$ is the Newtonian gravitational potential
\begin{equation}\label{eqn:Phieqn}
\Phi({\bf r},t)=-G\int d^3 r^\prime\frac{\rho({\bf r}^\prime,t)}{|{\bf r}-{\bf r}^\prime|}.
\end{equation}
Hence the  $\Phi$  field acts as the source term for  the velocity potential. Note that in the Newtonian
theory of gravity one has the choice of using either the acceleration field ${\bf g}$ or the velocity field
${\bf v}$. However in the new theory of gravity this choice is no longer available: the fundamental
dynamical degree of freedom is necessarily the ${\bf v}$ field, again because of the presence of the $C({\bf
v})$ term, which obviously cannot be written in terms of ${\bf g}$.  If we were to ignore time-dependent behaviour
(\ref{eqn:ueqn2}) gives
\begin{equation}\label{eqn:veqnagain}
|{\bf v}({\bf r})|^2=\frac{2}{4\pi}\int d^3
r^\prime\frac{C({\bf v}({\bf r}^\prime))}{|{\bf r}-{\bf r}^\prime|}-2\Phi({\bf r}).
\end{equation}
This non-linear equation clearly cannot be solved for ${\bf v}({\bf r})$ as its direction is not
specified.  This form makes it clear that we should expect gravitational waves, but certainly not waves travelling at
the speed of light as $c$ does not appear in (\ref{eqn:ueqn2}). Note that (\ref{eqn:ueqn2}) involves
`action-at-a-distance' effects, as there is no time-delay in the denominators. This was a feature of Newton's
original theory of gravity.  Here it is understood to be caused by the underlying quantum-foam dynamics (QHFT) which
reaches this classical `flow' description by ongoing non-local and instantaneous wavefunctional collapses, as
discussed in
\cite{NovaBook}. Contrary to popular belief even GR has this `action-at-a-distance' feature, as the reformulation of
GR via the Panlev\'{e}-Gullstrand metric  leads also  to an equation of the form in   (\ref{eqn:ueqn}), but
with the $C({\bf v})$ term absent.

\section[ Gravitational Waves]{ Gravitational Waves \label{section:waves}}
Newtonian gravity in its original `force' formalism (\ref{eqn:g1}) does not admit any wave phenomena.
However the completely equivalent `in-flow' formalism in (\ref{eqn:f2}) and (\ref{eqn:f3}) does admit  wave
phenomena. For the simpler case of zero vorticity, and so permitting the velocity potential
description, and also neglecting the `dark matter' term $C({\bf v})$, 
then (\ref{eqn:ueqn2}) becomes \index{velocity potential}
\begin{equation}
\frac{\partial  u}{\partial t}+\frac{1}{2}(\nabla u)^2=-\Phi.
\label{eqn:NGu}\end{equation}
and 
\begin{equation}{\bf g}=\frac{ \partial \nabla u}{\partial t}+\frac{1}{2}\nabla(\nabla u)^2, 
\end{equation} 
which
together reproduce (\ref{eqn:gPhi}), even when the flow is time-dependent. Suppose that (\ref{eqn:NGu}) has 
for a static matter density a static solution $u_0({\bf r})$ with corresponding velocity field ${\bf
v}_0({\bf r})$, and with corresponding acceleration
${\bf g}_0(\bf r)$. Then we look for time dependent perturbative solutions of (\ref{eqn:NGu}) with
$u=u_0+ \overline{u}$. To first order in $\overline{u}$ we then have
\begin{equation}
\frac{\partial {\overline u({\bf r},t)}}{\partial t}=-{\bf \nabla}\overline{u}({\bf r},t).{\bf
\nabla}u_0({\bf r}).
\label{eqn:ueqn3}
\end{equation}
This equation then has wave solutions of the form $\overline{u}({\bf r},t)=A\cos({\bf k}.{\bf r}-\omega t)$
where $\omega({\bf k},{\bf r})={\bf v}_0({\bf r}).{\bf k}$, for wavelengths short compared to 
the scale of changes in  ${\bf v}_0({\bf r})$. The phase velocity of these waves is then
${\bf v}_\phi={\bf v}_0$, and the group velocity is ${\bf v}_g={\bf \nabla}_k\omega={\bf 
v}_0$. Then the velocity field is 
\begin{equation}
{\bf v}({\bf r},t)={\bf v}_0({\bf r})-A{\bf k}\sin({\bf k}.{\bf r}-w({\bf k},{\bf r})t).
\end{equation}

But are these wave solutions physical\index{Newtonian gravitational waves}, or are they a mere artifact of
the in-flow formalism?  First note that the wave phenomena do not cause any gravitational effects, because
the acceleration field is independent of their existence;  whether they are present or not does not affect
 ${\bf g}({\bf r})$.  This question is equivalent to asking which of the fields ${\bf v}$ or ${\bf g}$ is
the fundamental quantity.  As we have already noted  the velocity field ${\bf v}$  and
these wave phenomena have already  been observed \cite{NovaBook,RGC,AMGE}. Indeed it is even possible that the
effects of such waves are present in the Michelson-Morley 1887 fringe shift data.  This would imply that the real
gravitational waves have actually been observed for over 100 years. 

Within the new theory of gravity these
waves do affect the acceleration field ${\bf g}$, via the new $C({\bf v})$ term. Numerical studies have shown these
wave effects, and that even when the `dark matter' effect is retained  this wave  phenomena persists.  The
observational evidence is that these gravitational waves are apparently present in the Milky Way and local galactic
cluster, as revealed in the analysis of data from at least three distinct observations of absolute motion effects
\cite{AMGE}.

\section[ Frame-Dragging Effect as an In-Flow Vorticity  Effect]{ Frame-Dragging Effect as an In-Flow Vorticity 
Effect \label{section:lense}}

Here we briefly note that (\ref{eqn:BS2}) and  the vorticity dependent term in (\ref{eqn:G7}) together explain the
frame-dragging  effect.  For the case where ${\bf v}_R$ is
determined solely by the rotation of the earth (\ref{eqn:BS2}) gives, outside of the earth, the dipole form
\begin{equation}
\omega({\bf r})=-4\frac{G}{c^2}\frac{3({\bf r}.{\bf L}){\bf r}-r^2{\bf L}}{2 r^5},
\label{eqn:earthvorticity}\end{equation}
where ${\bf L}$ is the angular momentum of the earth, and ${\bf r}$ is the distance from the centre of the earth.
Here spherical symmetry of the earth is assumed. When used in (\ref{eqn:G7}) the precession of a spinning  sphere,
caused by the $\omega\times {\bf v}_R$ term where here ${\bf v}_R$ is used to describe the rotation of the sphere,
may be used to detect the vorticity in (\ref{eqn:earthvorticity}), as in the Gravity  Probe B. This effect
has always caused interpretational problems in General Relativity: what system is it that acts as a frame
of reference in defining the rotation of the earth? Answers usually invoked some Machian explanation,
namely that the rotation was defined relative to the universe as a whole. In the  velocity-field 
formalism for gravity the rotation is relative to the local quantum-foam substratum, and the rotation of
the earth is affecting the in-flow component to the extent that it slightly drags the in-flow, that is,
it imparts some of its rotation to the in-flow, as described by the above  vorticity. In GR the
vorticity  field $\omega$ is known as the `gravitomagnetic' field, because of its role in the
Lorentz-like velocity-dependent acceleration in (\ref{eqn:G7}). The expression for the vorticity in 
(\ref{eqn:BS2}) would also appear to have contributions from the absolute linear motion of the earth. 
Such  an effect can be tested by the Gravity  Probe B as the  test-sphere spin-precession  would then be 
different in both magnitude and direction. 

\section{ Gravitational Anomalies}\label{section:gravitationalanomalies}

There are numerous  gravitational anomalies\index{gravitational anomalies}, including not  only the
spiral-galaxy `dark matter' effect and problems in measuring
$G$, but as well there  are others that   are not well-known in physics, presumably because their existence is
incompatible with the Newtonian or the  Hilbert-Einstein gravity theories.

The most significant of these anomalies is the Allais effect \cite{Allais}. In the 1950's  Allais conducted a long
series of experiments using a paraconical pendulum, which can be thought of as a  Foucault pendulum with a short
arm length and a special pivot mechanism. These observations revealed pendulum precession effects that are
distinct from the Foucault pendulum precession, which mainly manifests in the case of a very long pendulum,
associated with the position of the moon, but with a magnitude  very much larger than the well-known tidal effects.   
However in June 1954 Allais reported that the
paraconical pendulum exhibited peculiar movements at the time of a solar eclipse.  Allais was recording the
precession of the pendulum in Paris. Coincidentally during the 30 day observation period a partial solar
eclipse occurred at Paris on June 30.  During the eclipse the precession of the pendulum was seen to be disturbed. 
Similar results were obtained during another solar eclipse on October 29 1959.  There have been other repeats of the
Allais experiment with varying results.  

Another anomaly was reported by Saxl and Allen \cite{Saxl} during the solar eclipse of March 7 1970.  Significant
variations in the period of a torsional pendulum were observed  both during the eclipse and as well in the hours just
preceding and just following the eclipse.  The effects seem  to suggest that an ``apparent wavelike structure has been observed over the
course of many years at our Harvard laboratory'', where the wavelike structure is present and reproducible even in the
absence of an eclipse. 

Again Zhou and Huang \cite{Zhou}  report various clock anomalies occurring during the solar eclipses of September 23
1987,  March 18 1988 and  July 22 1990 observed using atomic clocks.

Another anomaly is of course the `dark matter' effect associated  with the rotational velocities of objects in spiral
galaxies. This anomaly
led to the introduction of the `dark matter' concept - but with no such matter ever having been detected, despite
extensive searches. This anomaly was compounded when recently observations of the rotational velocities of objects
within elliptical galaxies was seen to require very little `dark matter'.  Of course this is a simple consequence of
the new theory of gravity, as we shall see.    

All these anomalies, including the $g$ anomaly in Sect.\ref{section:theborehole} and others such as the
 the solar neutrino flux deficiency problem were clearly indicating that
  gravity has aspects to it that are not within the prevailing theories.

\section[ The Borehole $g$ Anomaly and the Fine Structure Constant]{  The  Borehole $g$ Anomaly and the Fine Structure
\newline \mbox{\  }Constant\label{section:theborehole}}

\begin{figure}[ht]
\hspace{40mm}\includegraphics[scale=0.9]{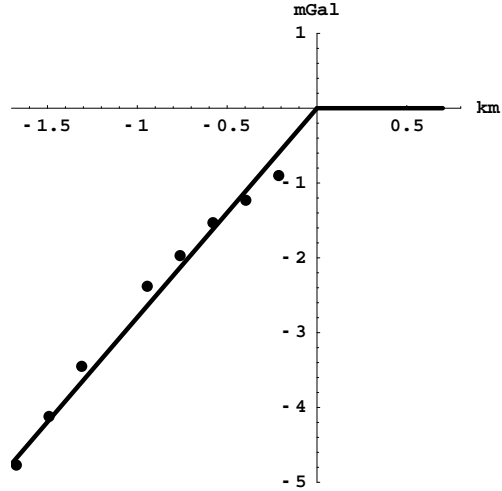}
\caption{\small{ The data shows the gravity residuals for the Greenland Ice Cap \cite{Greenland} Airy measurements of the
$g(r)$  profile,  defined as
$\Delta g(r) = g_{Newton}-g_{observed}$, and measured in mGal (1mGal $ =10^{-3}$ cm/s$^2$), plotted against depth in km.  Using
(\ref{eqn:BHdeltag2}) we obtain $\alpha^{-1}=139 \pm  5 $ from fitting the slope of the data, as shown.}
\label{fig:Greenland}}\end{figure}

Stacey and others \cite{Stacey1,HST86,HT87,Stacey2} have found evidence for non-Newtonian gravitation from
gravimetric measurements (Airy experiments\index{Airy experiments}) in mines and boreholes\index{borehole $g$
anomaly}.  The discovery was that the measured value of $g$ down mines and boreholes became greater than that
predicted by the Newtonian theory, given the density profile
$\rho(r)$ implied by sampling, and so implying a defect in Newtonian gravity, as shown in Fig.\ref{fig:Greenland} for
the Greenland Ice shelf borehole measurements\index{Greenland  borehole measurements}.  The results were
interpreted and analysed using either a value of
$G$  different to but larger than that found in laboratory experiments or by assuming a short range Yukawa type
force in addition to the Newtonian `inverse-square law'.  Numerous experiments were carried out in which
$g$ was measured as a function of depth, and also as a function of height above ground level using towers. The
tower experiments \cite{Thomas89,Jekeli} did not indicate any non-Newtonian effect, and so implied that the
extra Yukawa force explanation was not viable.   The combined results appeared to have resulted in
confusion and eventually the experimental effect was dismissed as being caused by erroneous
density sampling.  However the new theory of gravity predicts such an effect, and in particular that the
effect should manifest within the earth but not above it, as was in fact observed. The effect predicted is that
$d\Delta g(r)/dr$ should be discontinuous at the boundary, as shown in Fig.\ref{fig:Greenland}.    Essentially
this effect is caused by the new
$C({\bf v})$ term in the in-flow theory of gravity.

When the matter density and the flow are both
spherically symmetric and stationary in time (\ref{eqn:f3extend}) becomes, with  $v^\prime \equiv dv/dr$,
\begin{equation}
2\frac{vv^\prime}{r} +(v^\prime)^2 + vv^{\prime\prime} =-4\pi G\rho(r)-4\pi G \rho_{DM}(r), 
\label{eqn:BHInFlowRadial}
\end{equation}
and then
\begin{equation}
\rho_{DM}(r)= \frac{\alpha}{32\pi G}\left(\frac{v^2}{2r^2}+ \frac{vv^\prime}{r}\right).
\label{eqn:BHdm1}\end{equation}
Eqn.(\ref{eqn:BHInFlowRadial}) may be written in a non-linear integral form
\begin{equation}
v^2(r)=\frac{8\pi G}{r}\int_0^r s^2 \left[\rho(s)+\rho_{DM}(s)\right]ds+8\pi G\int_r^\infty s
\left[\rho(s)+\rho_{DM}(s)\right]ds,
\label{eqn:BHintegralEqn}\end{equation}
which follows from  evaluating 
\begin{equation}\label{eqn:vvsp}
|{\bf v}({\bf r})|^2=2G\int d^3
r^\prime\frac{\rho(r^\prime)+\rho_{DM}(r^\prime)}{|{\bf r}-{\bf r}^\prime|}.
\end{equation}
in the case of spherical symmetry and a radial in-flow.

First  consider solutions to (\ref{eqn:BHintegralEqn}) in the perturbative regime. Iterating once we find 
\begin{equation}
\rho_{DM}(r)=\frac{\alpha}{2r^2}\int_r^\infty s\rho(s)ds+O(\alpha^2),
\label{eqn:perturbative}\end{equation}
so that in spherical systems the `dark matter' effect is concentrated near the centre, and we find that the total `dark
matter'
\begin{equation}
M_{DM}\equiv 4\pi\int_0^\infty r^2\rho_{DM}(r)dr=\frac{4\pi\alpha}{2}\int_0^\infty
r^2\rho(r)dr+O(\alpha^2)=\frac{\alpha}{2}M+O(\alpha^2),
\label{eqn:BHTotalDM}\end{equation}
where $M$ is the total amount of (actual) matter. Hence to $O(\alpha)$   $M_{DM}/M=\alpha/2$ independently of the matter
density profile.

 When the matter density $\rho(r)=0$ for $r\geq R$, as for the earth, then
 we also obtain, to $O(\alpha)$, from
(\ref{eqn:f2}) and (\ref{eqn:BHintegralEqn}) Newton's `inverse square law' for $r > R$
\begin{equation}
g(r)=\left\{ \begin{tabular}{ l} 
$\displaystyle{-\frac{(1+\displaystyle{\frac{\alpha}{2}}) GM}{r^2},\mbox{\ \ } r > R,}$  \\  
$\displaystyle{-\frac{4\pi G}{r^2}\int_0^rs^2\rho(s) ds
-\frac{2\pi\alpha G}{r^2}\int_0^r\left(\int_s^R s^\prime\rho(s^\prime) ds^\prime\right) ds,\mbox{\ \ } r < R,} $\\ 
\end{tabular}\right.   
\label{eqn:BHISL2}\end{equation}
and we see that the effective Newtonian gravitational constant in (\ref{eqn:BHISL2}) is $G_N=(1+\frac{\alpha}{2})G$ which is 
different to the fundamental gravitational constant
$G$ in (\ref{eqn:f3extend}). The result in (\ref{eqn:BHISL2}), which is different from that of the Newtonian theory 
($\alpha=0$) has actually been observed in mine/borehole measurements \cite{Stacey1,HST86,Greenland} of
$g(r)$, though of course there had been no explanation for the effect, and indeed the
reality of the effect was eventually doubted.    
The gravity residual \cite{Stacey1,HST86,Greenland}  is defined as
\begin{eqnarray}
\Delta g(r)&\equiv & g(r)_{Newton}-g(r)_{observed}\\ 
&=&g(r)_{Newton}-g(r). 
\label{eqn:BHdeltag1}\end{eqnarray}
The `Newtonian theory' assumed in the determination of the gravity residuals  is, in the present context,
\begin{equation}
g(r)_{Newton}=\left\{ \begin{tabular}{ l} 
$\displaystyle{-\frac{G_N M}{r^2},\mbox{\ \ } r > R,}$  \\  
$\displaystyle{-\frac{4\pi G_N}{r^2}\int_0^rs^2\rho(s) ds,\mbox{\ \ } r < R,} $\\ 
\end{tabular}\right.   
\label{eqn:BHearthg2}\end{equation}
with $G_N=(1+\frac{\alpha}{2})G$. Then $\Delta g(r)$ is found to be, to 1st order in $\alpha$ and in $R-r$,  i.e.
near the surface, 
\begin{equation}
\Delta g(r)=\left\{ \begin{tabular}{ l} 
$\displaystyle{\mbox{\ \ }0, \mbox{\ \ } r> R,}$  \\   
$\displaystyle{-2\pi\alpha G_N\rho(R)(R-r),\mbox{\ \ } r < R.} $\\ 
\end{tabular}\right.   
\label{eqn:BHdeltag2}\end{equation}
which is the form actually observed  \cite{Stacey1,HST86,Greenland}. 
So outside of the spherical earth the Newtonian theory and the in-flow theory are
indistinguishable, as indicated by the horizontal line, for $r>R$, in Fig.\ref{fig:Greenland}.  However inside the earth the
two theories give a different dependence on $r$, due to the `dark matter' effect within the earth.

  Gravity  residuals\index{gravity residuals}  from a borehole
into the Greenland Ice Cap  were determined  down to a depth of 1.5km \cite{Greenland}. The ice had a density of
$\rho(R)=930$ kg/m$^3$, and from (\ref{eqn:BHdeltag2}), using $G_N=6.6742\times10^{-11}$ m$^3$s$^{-2}$kg$^{-1}$, we obtain from a
linear fit to the slope of the data points in Fig.\ref{fig:Greenland} that
$\alpha^{-1}=139\pm 5$, which equals the value of the fine structure constant  $\alpha^{-1}=137.036$ to within the errors, and
for this reason we identify the $\alpha$ constant in (\ref{eqn:f3extend}) as being the fine structure constant.

The so called fine structure constant\index{fine structure constant ($\alpha$)} $\alpha$ was introduced into
physics by Sommerfeld  in 1916. Sommerfeld extended the Bohr theory of atoms to include elliptical orbits and the
relativistic dependence of mass on speed.  The result for
  a typical energy difference is
 \begin{equation}
E_{nk}=-\frac{mc^2\alpha^2 k}{2n^2}\left( 1+\frac{\alpha^2}{n^2}\left(\frac{n}{k}-\frac{3}{4} \right) \right).
\label{eqn:fine}\end{equation}
We see that the leading term contains $\alpha^2$, as well as the second term which introduces another
$\alpha^2$. It is because of its presence in this second order  term that $\alpha$ is called the fine
structure constant, though that is really a misnomer, for $\alpha$ determines also the Bohr energies, as is
seen once we compare the atomic energy levels with the rest mass energy of the electron, as in
(\ref{eqn:fine}). The occurence of $\alpha$ in the self-interaction dynamics of space implies that the
stochastic processing in the information-theoretic  {\it process physics}, see \cite{NovaBook}, involves a probability
measure that not only manifests in this spatial dynamics but also manifests in  Quantum Electrodynamics, where there
$\alpha$ is a measure of the probablity for a charged particle to emit or absorb a photon. Clearly we are seeing
evidence of a deep unification of fundamental physics. 

\section[ Measurements of $G$]{  Measurements of $G$ and the Fine Structure
\newline \mbox{\  }Constant\label{section:measurementsofG}}

As already noted  Newton's Inverse Square Law\index{inverse square law} of
Gravitation may only be strictly valid in special cases.  The  theory that
gravitational effects arise from inhomogeneities in the quantum foam flow  implies that there is no
`universal law of gravitation' because the inhomogeneities are determined by non-linear `fluid equations'
and the solutions  have no form which could be described by a `universal law'.  Fundamentally there is no
generic fluid flow behaviour. The Inverse Square Law is then only an approximation, with  large
deviations seen in the case of spiral galaxies. Nevertheless Newton's gravitational constant $G$
will have a definite value as it quantifies the effective rate at which matter dissipates the
information content of space.  
\begin{figure}
\hspace{15mm}\includegraphics[scale=1.2]{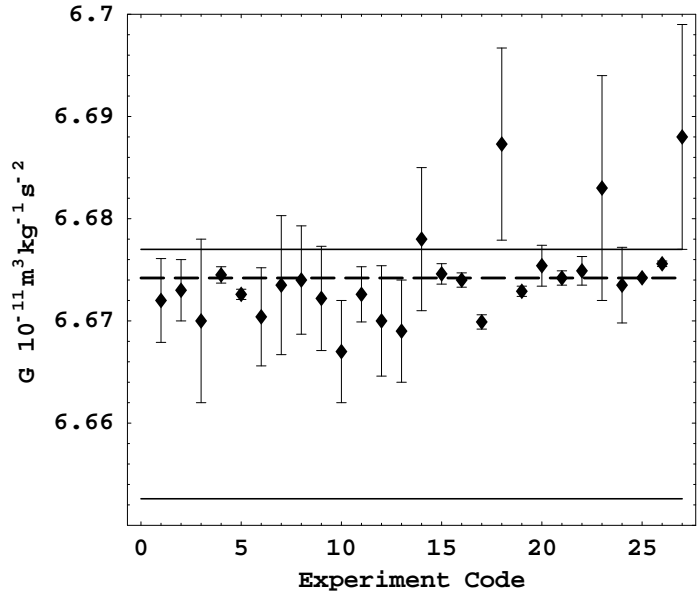}
\caption{\small{Results of precision measurements of $G$ published in the 
last sixty years in which the Newtonian theory was used to analyse the data.  These results 
show  the presence of
a  systematic effect, not in the Newtonian theory, of fractional size $\Delta G/G \approx \alpha/4$. 
The upper horizontal  line shows the value of $G$ from ocean Airy measurements \cite{Ocean}, while the dotted
 line shows the current CODATA
$G$ value. The lower horizontal line shows
the value of $G$ after removing the `dark matter' effects from the \cite{Ocean} $G$ value. \newline {\bf
Experiment Codes:} {\bf 1:}  Gaithersburg 1942
\cite{Gaithersburg42}, {\bf 2:}  Magny-les-Hameaux 1971 \cite{Magny-les-Hameaux}, 
{\bf 3:}  Budapest 1974   \cite{Budapest}, 
{\bf 4;}  Moscow 1979  \cite{Moscow79},
{\bf 5:}  Gaithersburg 1982 \cite{Gaithersburg82}, 
{\bf 6-19:}  Fribourg   Oct 84, Nov 84, Dec 84, Feb 85 \cite{Fribourg},
{\bf 10:}    Braunschweig 1987   \cite{Braunschweig87},
{\bf 11:}    Dye 3 Greenland   1995       \cite{Dye3},
{\bf 12:}    Gigerwald Lake 1994   \cite{Gigerwaldlake94}, 
{\bf 13-14:} Gigerwald lake19 95  112m, 88m     \cite{Gigerwaldlake95},
{\bf 15:}     Lower Hutt 1995    MSL  \cite{LowerHutt95}, 
{\bf 16:}  Los Alamos 1997 \cite{Los Alamos}, 
{\bf 17:}    Wuhan 1998  \cite{Wuhan},
{\bf 18:}    Boulder JILA 1998  \cite{Boulder}, 
{\bf 19:}    Moscow 1998  \cite{Moscow98}, 
{\bf 20:}    Zurich 1998  \cite{Zurich98}, 
{\bf 21:}     Lower Hutt MSL 1999   \cite{LowerHutt99}, 
{\bf 22:}    Zurich 1999   \cite{Zurich99}, 
{\bf 23:}    Sevres 1999  \cite{Sevres99}, 
{\bf 24:}   Wuppertal 1999  \cite{Wuppertal},
{\bf 25:}     Seattle 2000 \cite{Seattle},
{\bf 26:}     Sevres 2001  \cite{Sevres01}, 
{\bf 27:}     Lake Brasimone 2001 \cite{lake Brasimone}. 
  }   
\label{fig:GData}}\end{figure}

From these considerations it
follows that the measurement of the value of $G$ will be difficult as the measurement\index{measurement of $G$} of
the  forces between two of more objects, which is the usual method of measuring $G$, will depend on the geometry
of the spatial positioning of these objects  in a way not previously accounted for because the
Newtonian Inverse Square Law has always been assumed, or in some case  a specified change in the
form of the law has been used.  But in all cases a `law' has been assumed, and this may have been
the flaw in  the analysis of data from such experiments.  This implies that the value of
$G$ from such experiments will show some variability as a systematic effect has  been neglected in
analysing the experimental data.  So
experimental measurements of $G$  should show an unexpected contextuality.  As well the influence of
surrounding matter has also not been properly accounted for. Of course  any effects of turbulence in
the inhomogeneities of the flow has presumably also never even been contemplated.  
The first measurement of $G$ was in 1798 by Cavendish using a torsional balance.  As the precision of
experiments increased over the years and a variety of techniques used  the disparity between the
values of $G$ has actually increased \cite{Gillies}.  Fig.\ref{fig:GData} shows the results from precision 
measurements of
$G$ over the last 60 years. As can be seen  one indication of the contextuality is that measurements of $G$  produce
values that differ by nearly 40 times their individual error estimates. In 1998 CODATA increased the uncertainty in
$G$, shown by the dotted line in \ref{fig:GData}, from 0.013\% to 0.15\%.

 Note that the relative spread $\Delta G_N/G_N \approx O(\alpha)$, as we would now expect.
Essentially the different Cavendish-type laboratory experiments used different matter geometries and, as we have
seen, the
 geometry of the masses  has an  effect on the in-flow, and so on the measured force between the masses.  Only
for the borehole-type experiments do we have a complete analytic analysis, in Sect.\ref{section:theborehole}, and
an ocean measurement is of that type, and experiment
\cite{Ocean}  gives a $G_N=(6.677\pm 0.013)\times 10^{-11} $ m$^2$s$^{-2}$kg$^{-1}$, shown by the upper
horizontal line in Fig.\ref{fig:GData}.
 From that value  
 we may extract the value of the `fundamental gravitational constant' $G$ by removing the `dark matter' effect:
$G \approx(1-\frac{\alpha}{2})G_N= (6.6526 \pm 0.013)\times 10^{-11} $ m$^2$s$^{-2}$kg$^{-1}$, shown by the lower
 horizontal line in Fig.\ref{fig:GData},
compared to the current CODATA value of  
$G_N=(6.6742 \pm 0.001)\times 10^{-11}$ m$^2$s$^{-2}$kg$^{-1}$, which is contaminated with   `dark matter' effects. 
Then in the various experiments, without explicitly
computing the `dark matter' effect, one will find an `effective' value of $G_N>G$ that depends on the geometry of the masses.
A re-analysis of the data in  Fig.\ref{fig:GData} using the in-flow theory is predicted to resolve these apparent
discrepancies.  
 Examples of  how the quantum-foam fluid-flow
theory of gravity alters the analysis of data from  Cavendish-type experiments are given in 
Sect.\ref{section:laboratory}, and in general the `dark matter' effects are of order $\alpha$.

\section[ Gravitational Attractors - New Black Holes]{   Gravitational Attractors - New Black Holes
\label{section:gravitationalattractors}}

Here we consider  a new phenomena which is not in either the Newtonian or Einsteinian theories of gravity, namely
the existence of   gravitational attractors\index{gravitational attractor}.  Such an attractor  may exist by
itself or it may be accompanied by matter, as in the case of planets, stars, globular clusters and galaxies, both
elliptical and spiral. As we have seen in Sect.\ref{section:theborehole} the existence of such an  attractor at the
centre of the earth is suggested by the borehole $g$ anomaly data.  Here we develop the general
theory of these attractors.  Indeed they are apparently a common occurrence.  Up to now the effects of these 
`attractors' in globular clusters\index{globular clusters}, quasistellar objects (QSO)\index{quasistellar objects
(QSO)}  and galaxies have been interpreted by astronomers as general relativity `black holes',\index{black hole} but
only by default as no other phenomenon was until now known which could account for the strong gravitational effects
observed at the centres of these systems. These attractors are self-sustaining quantum foam in-flows, and their
behaviour  is determined solely by the fine structure constant: they are quantum foam in-flow singularities where the
quantum-foam is destroyed, together with any matter that happnes to in-fall. So to that extent they are classical
manifestations of quantum gravity.  These attractors  have an event horizon\index{event horizon} where the in-flow
speed reaches the speed of light, and within this horizon the speed increases without limit to an  infinite speed at a
singular point. In-falling matter  can produce radiation from the heating effects associated with this in-fall. 

  However the existence of these in-flow singularities does
not require that they be formed by the collapse of matter, and they need not have  matter at their centres, so in
many respects they differ from the `black holes' of general relativity.  They also differ from these  `black holes' in
that their gravitational acceleration $g(r)$ is not given by Newton's inverse square law.  It is suggested here that
 along with matter and radiation, that  they formed the third component of the universe, apart from space itself. 
And that  they played a key role in the formation of certain gravitationally collapsed systems, such as spiral galaxies,
and that the long range nature of their acceleration field $g(r)$ explains the apparent relatively rapid formation
of  such structures in the early universe.

Here we first consider the special case of a  one-parameter class of matter-free spherical
attractors\index{spherical attractor}, and then the case  when the attractor is associated with matter. In this case
the attractor may be minimal\index{minimal attractor}  or non-minimal\index{non-minimal attractor}, with the
distinction determined by whether or not the attractor produces a long-range acceleration field, as for the
non-minimal attractors. We also consider  a class of non-spherical attractors, but the non-sphericity  produce
only short range effects.  For the case of spherical attractors we  determine the size of the event horizon. For
globular clusters  we can then predict the minimum mass of the  central attractor and compare that with the total
mass of the cluster. This ratio is shown to be equal to
$\alpha/2$, and this prediction is in agreement with the observations of the M15 and G1 globular clusters.  Hence
the globular clusters supply a striking confirmation of the new theory of gravity and its attractors.  In
Sect.\ref{section:laboratory} we show that these attractors may be experimentally studied in Cavendish-style
laboratory gravitational experiments, and so provide the opportunity for the first laboratory quantum gravity
experiments and indeed laboratory black hole experiments.

 There will be at least a minimal  attractor within the sun which can also be detected by
analysing the neutrino flux and its energy spectrum, as these attractors produce central gravitational forces very
different from the Newtonian theory, which is an essential input into current stellar dynamics.

\section[ Spherical  Gravitational Attractors]{ Spherical  Gravitational
Attractors\label{section:the gravitationalattractor}}

Here we reveal the one-parameter class of spherical attractors in the absence of matter. For a spherically
symmetric  in-flow $v(r,t)$ the basic in-flow equation (\ref{eqn:f3extend})  has the form
\begin{equation}
\frac{\partial v^\prime}{\partial t}+vv^{\prime\prime}+\frac{vv^\prime}{r} +(v^\prime)^2 
+\frac{\alpha}{2}\left(\frac{v^2}{2r^2}+
\frac{vv^\prime}{r}\right)=0, 
\label{eqn:InFlowRadialB}
\end{equation}
where $v^\prime=\partial v(r)/\partial r$.  For a stationary flow this equation becomes linear in $f(r)$ where 
$v(r)=\surd f(r)$ 
\begin{equation}
\frac{f^{\prime\prime}}{2}+\frac{f^\prime}{r}+\frac{\alpha}{2}(\frac{f}{2r^2}+
\frac{f^\prime}{2r})=0.
\label{eqn:feqn}\end{equation}
 The general solution of this homogeneous equation is 
\begin{equation}
f(r)=\frac{K}{r}+\frac{\beta}{r^{\alpha/2}},
\label{eqn:gensoln}\end{equation}
where  $K$ and  $\beta$ are arbitrary constants.
The effective `dark matter' density (\ref{eqn:BHdm1}) is then
\begin{equation}
\rho_{DM}(r)=\frac{\alpha\beta}{16\pi G}(1-\frac{\alpha}{2})\frac{1}{r^{2+\alpha/2}},
\label{eqn:DMdensity}\end{equation}
which essentially has the $1/r^2$  dependence, as seen in spiral galaxies and discussed later. Note that the   $K$
term does not contribute to $\rho_{DM}$. However the
$K$ term is not a solution of (\ref{eqn:f3extend}) for a stationary flow. The reason for this is somewhat subtle. 
By direct computation we would appear to obtain  that 
\begin{equation}\nabla^2\frac{1}{|{\bf
r}|}=\frac{1}{2}\frac{d^2}{dr^2}\frac{1}{r}+\frac{1}{r}\frac{d}{dr}\frac{1}{r}=0,
\end{equation} 
which is used in finding the $K$ term part of (\ref{eqn:gensoln}). But in fact the correct result is   
\begin{equation}
\nabla^2\frac{1}{|{\bf r}|}=-4\pi\delta^{(3)}({\bf r}).
\label{eqn:delta}\end{equation} 
This is confirmed
by applying the divergence theorem
\begin{equation}
\int d V {\bf \nabla}.{\bf w}=\int d {\bf A}.{\bf w},
\label{eqn:divergence}\end{equation} 
with ${\bf w}={\bf \nabla}(1/|{\bf r}|)$ for a spherical region. The RHS of
(\ref{eqn:divergence}) gives $-4\pi$ independent of the radius of the sphere. So  the LHS 
must have a delta-function distribution at ${\bf r}={\bf 0}$, as in (\ref{eqn:delta}).  Hence the $K$
term should not be present in (\ref{eqn:gensoln}). Essentially for this term to appear the RHS of 
(\ref{eqn:f3extend}) would have to have a  $-4\pi\delta^{(3)}({\bf r})$ term corresponding to a
point mass.  However by definition all of the matter density is included in $\rho({\bf r})$, and in the present case
there is no matter present at all. However for the $\beta$ term the result is different. By direct computation we
find that
\begin{equation} 
\nabla^2\frac{1}{|{\bf r}|^\alpha}=-\frac{\alpha(1-\alpha)}{r^{2+2\alpha}}.
\label{eqn:correct}\end{equation}
Using the divergence theorem again but now with ${\bf w}={\bf \nabla}(1/|{\bf r}|^\alpha)$ we find that
(\ref{eqn:divergence}) is satisfied, and so no delta-function distribution is needed on the RHS of  
(\ref{eqn:correct}). Hence the correct general solution of (\ref{eqn:feqn}) is
\begin{equation}
f(r)=\frac{\beta}{r^{\alpha/2}},
\label{eqn:gensolnb}\end{equation}
which defines a one-parameter class of spherically symmetric attractors. The gravitational acceleration produced by
this in-flow is
\begin{equation}
g(r)=\frac{1}{2}\frac{d f(r)}{dr}=-\frac{\alpha\beta}{4r^{1+\alpha/2}},
\label{eqn:gattractor}\end{equation}
which decreases  slowly with distance, compared to Newton's inverse square law.  Because these attractors can
be independent of matter  they would have arisen in the early universe during the formation of space itself. Once
matter had cooled to the recombination temperature of about 3000$^o$K, these attractors would have played a key role
in the formation of the first stars and the galaxies.  This attractor has a spatial in-flow with a speed
singularity\index{in-flow singularity} at $r=0$.  There is a spherical event  horizon, where $v=c$, at
$r_H=(\frac{\beta}{c^2})^{2/\alpha}$. Hence the attractor acts as a `black hole', but very much unlike the `black
hole' in general relativity. The  universe would have had these attractors as  primordial black holes  from the very
beginning.

\section[ Minimal  Attractor]{ Minimal  Attractor for a Uniform Density 
\newline \mbox{\  }Sphere\label{section:minimal}} 

Now consider the gravitational attractors that are formed by the presence of matter. 
Here the quantum foam in-flow associated with the matter appears to trigger a non-Newtonian in-flow
at the centre of the matter distribution.  Now for a spherically symmetric matter density and a 
spherically symmetric in-flow $v(r,t)$ the basic in-flow equation (\ref{eqn:f3extend})  has the
form\index{spherical attractor}
\begin{equation}
\frac{\partial v^\prime}{\partial t}+vv^{\prime\prime}+\frac{vv^\prime}{r} +(v^\prime)^2 
+\frac{\alpha}{2}\left(\frac{v^2}{2r^2}+
\frac{vv^\prime}{r}\right)=-4\pi G\rho(r), 
\label{eqn:InFlowRadialMinimal}
\end{equation}
Again for a stationary in-flow this equation becomes linear in $f(r)$ where  $v(r)=\surd f(r)$ 
\begin{equation}
\frac{f^{\prime\prime}}{2}+\frac{f^\prime}{r}+\frac{\alpha}{2}(\frac{f}{2r^2}+
\frac{f^\prime}{2r})=-4\pi
G\rho(r).
\label{eqn:feqnMimimal}\end{equation}
 Define the particular `matter dependent' solution of this
inhomogeneous equation  to be  
$f_m(r)$. 
Then the general solution of (\ref{eqn:InFlowRadialB}) is the sum of this particular solution and the
solutions of the homogeneous equation,
\begin{equation}
f(r)=\frac{\beta}{r^{\alpha/2}}+f_m(r),
\label{eqn:gensolnMimimal}\end{equation}
where   $\beta$ is again an arbitrary constant.
The effective `dark matter' density (\ref{eqn:BHdm1}) is now
\begin{equation}
\rho_{DM}(r)=\frac{\alpha\beta}{16\pi G}(1-\frac{\alpha}{2})\frac{1}{r^{2+\alpha/2}}+
\frac{\alpha}{2}(\frac{f_m}{2r^2}+\frac{f^\prime_m}{2r}).
\label{eqn:DMdensityMimimal}\end{equation}

Let us now consider the solution of (\ref{eqn:feqnMimimal})  for a piece-wise constant matter density,
in particular for a sphere of radius $R$ of uniform density $\rho$: 
 \begin{equation}\rho(r)=\rho,\mbox{ \ }
0<r<R;\mbox{ \ } \rho(r)=0,\mbox{ \ } r>R.
\label{eqn:uniformsphere}\end{equation}
The  solution of   (\ref{eqn:feqnMimimal}) is then found to be, in each region,
\begin{equation}
f(r)=\left\{ \begin{tabular}{ l} 
$\displaystyle{\frac{\beta}{r^{\alpha/2}}-
\frac{16\pi\rho G r^2}{3(4+\alpha)},\mbox{\ \ } 0< r < R,}$
\\  
$\displaystyle{\frac{K}{r}+\frac{\overline{\beta}}{r^{\alpha/2}},\mbox{\ \ } r > R, }$\\ 
\end{tabular}\right.   
\label{eqn:solidsphere}\end{equation}
where in general $\beta$ and $\overline{\beta}$ have different values. As well the $K$ term is permitted
in the external region. Indeed for a non-stepwise density the $K$ term would arise as the asymptotic
or the `matter dependent' solution $f_m(r)$.  The  complete solution is obtained by ensuring that 
$f(r)$ and $f^\prime(r)$ are continuous at $r=R$, which is required of the 2nd order differential
equation. Let us first consider the critical case where the gravitational attractor does not extend
beyond the sphere, i.e. $\overline{\beta}=0$.   This shall result in what is defined here to be a `minimal
attractor'. Then we find that
\begin{equation}
\beta=\frac{16\pi\rho G R^{2+\alpha/2}}{(1-\alpha/2)(4+\alpha)},
\label{eqn:betavalue}\end{equation}
\begin{equation}
K=16\pi \rho G
R^3\frac{2+\alpha/2}{3(1-\alpha/2)(4+\alpha)}=2(1+\frac{\alpha}{2}+O(\alpha^2)\mbox{\ })MG,
\label{eqn:KvalueMimimal}\end{equation}
where $M$ is the total matter content of the sphere.  Then for the `dark matter' density we obtain
\begin{equation}
\rho_{DM}(r)=\left\{ \begin{tabular}{ l} 
$\displaystyle{\frac{\alpha\beta}{16\pi
G}(1-\frac{\alpha}{2})\frac{1}{r^{2+\alpha/2}}-\frac{\alpha\rho}{4+\alpha}=\frac{\alpha\rho}{4}
\left(\left(\frac{R}{r}\right)^{2+\alpha/2}-1 \right),\mbox{\ \ } 0< r < R,}$ \\  
$0,\mbox {\ \  }  r > R.$\\ \end{tabular}\right.   
\label{eqn:DMsolidsphere}\end{equation}
which agrees with (\ref{eqn:perturbative}), to $O(\alpha)$, for a uniform matter density.

\begin{figure}
\hspace{0mm}\includegraphics[scale=1.8]{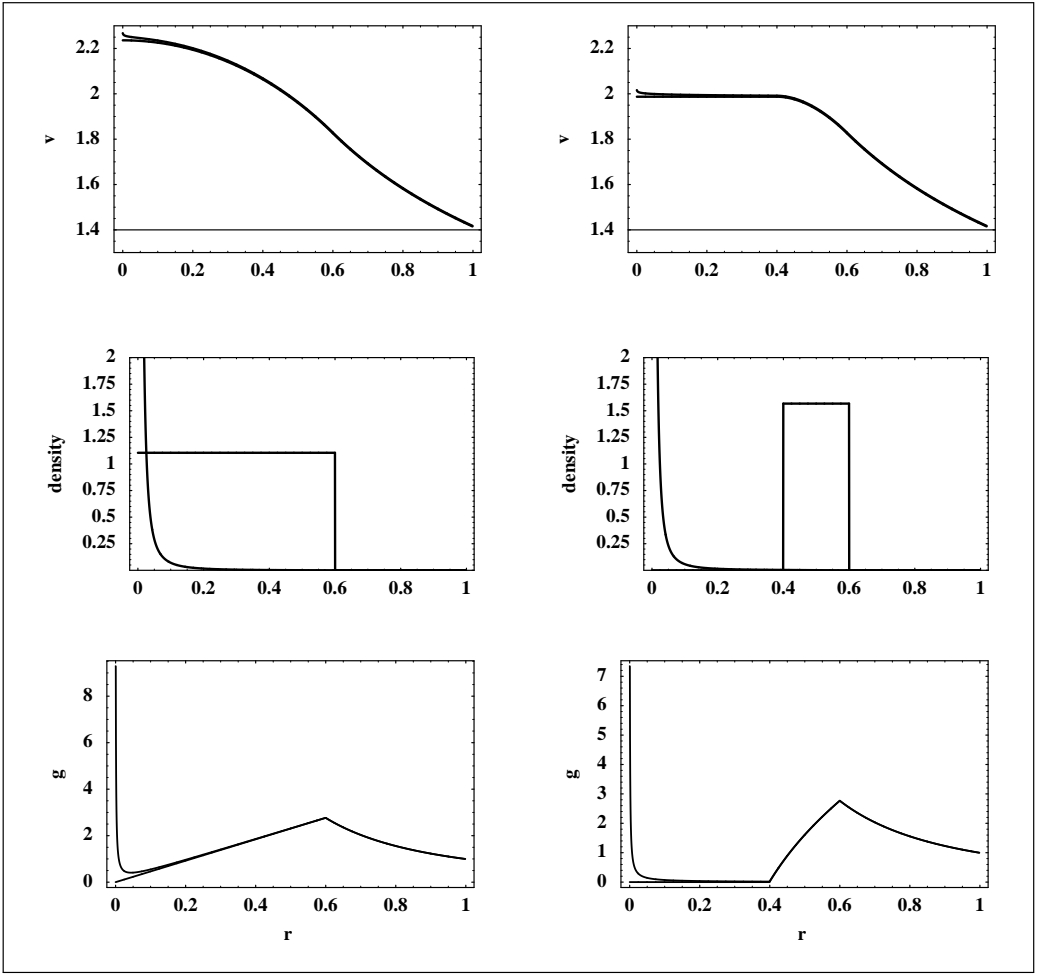}
\caption{{\small Solutions of (\ref{eqn:feqnMimimal}) for a uniform density sphere  (on the
left), and for  a spherical shell\index{spherical shell} (on the right). The upper plots show the in-flow speeds for
both Newtonian gravity and for the new theory of gravity, which displays the increase in speed near the attractor at
$r=0$.  The middle plots are the matter density profiles, with the `dark matter' density peaking
at $r=0$. The lower plots show the gravitational acceleration $|g(r)|$, with again a comparison of
the Newtonian gravity and the new theory, which shows large accelerations near the attractor at
$r=0$.}\label{fig:OneDPlots}}
\end{figure}
Eqn.(\ref{eqn:KvalueMimimal})  gives the external gravitational acceleration
\begin{equation}
g(r)=-\frac{\displaystyle{(1+\frac{\alpha}{2}+O(\alpha^2)\mbox{\ })}MG}{r^2},\mbox{\ \ }
r>R,
\label{eqn:externalg}\end{equation}
in agreement with (\ref{eqn:BHISL2}). So for this case the system would produce Keplerian orbits for  small
test objects in orbit about this sphere. So the perturbative analysis in Sect.\ref{section:theborehole} gave
rise to a minimal attractor. For such a uniform density sphere the in-flow speed $v(r)$, matter density
$\rho(r)$ and `dark matter' density  $\rho_{DM}(r)$, and the acceleration
$-g(r)$ are plotted in the left column of Fig.\ref{fig:OneDPlots}, including the special case $\alpha=0$
which gives the Newtonian gravity results.  Shown in  the right hand column of Fig.\ref{fig:OneDPlots}
are the corresponding results for a spherical shell with the same total mass $M$, as the sphere.  In
general the new theory predicts a gravitational attractor at the centre of all  matter
distributions.  For the case of a spherical system the effective mass of the attractor \index{attractor mass}is
$M_{DM}=\alpha M/2$, as measured by the  external gravitational acceleration in (\ref{eqn:externalg}).  Note that for
the minimal attractor the in-flow is uniquely determined  by the matter density, apart of course from time-dependent
behaviour. The minimal attractor is caused by the matter induced in-flow, and as such is not the result of a
primordial attractor.  
 
\section[ Non-Minimal  Attractor]{ Non-Minimal  Attractor for a Uniform Density 
\newline \mbox{\ }Sphere\label{section:nonminimal}} 

Now consider the more general case of a non-minimal attractor\index{non-minimal attractor} for which the
non-Newtonian acceleration field (\ref{eqn:gattractor})  extends beyond the matter density. These attractors have a
primordial origin, and would have played a critical role in the formation of the matter system from a gas cloud.
 In the
non-minimal case with $\overline{\beta} > 0$, we find solutions parametrised by $M$ and an
arbitrary valued $\overline{\beta}$.   Again for a sphere of uniform density  as in
(\ref{eqn:uniformsphere}) and with regional solutions as in  (\ref{eqn:solidsphere}), matching $f(r)$  and
$f^\prime(r)$ at $r=R$ gives
\begin{equation}
\beta=\overline{\beta}+\frac{16\pi\rho G R^{2+\alpha/2}}{(1-\alpha/2)(4+\alpha)},
\label{eqn:betabar}\end{equation}
\begin{equation}
K=16\pi \rho G
R^3\frac{2+\alpha/2}{3(1-\alpha/2)(4+\alpha)}=2(1+\frac{\alpha}{2}+O(\alpha^2)\mbox{\ })MG.
\label{eqn:KvalueNonMimimal}\end{equation}
Hence the value of $K$ is unchanged from the minimal case, while the value of $\beta$ is simply increased by
$\overline{\beta}$ from the minimal case.  Then  for the `dark matter' density we obtain
\begin{equation}
\rho_{DM}(r)=\left\{ \begin{tabular}{ l} 
$\displaystyle{\frac{\alpha\overline{\beta}}{16\pi
G}(1-\frac{\alpha}{2})\frac{1}{r^{2+\alpha/2}}+\frac{\alpha\rho}{4}
\left(\left(\frac{R}{r}\right)^{2+\alpha/2}-1\right)
,\mbox{\ \ } 0< r < R,}$ \\  
$\displaystyle{\frac{\alpha\overline{\beta}}{16\pi
G}(1-\frac{\alpha}{2})\frac{1}{r^{2+\alpha/2}}},\mbox {\ \  }  r > R.$\\ \end{tabular}\right.   
\label{eqn:DMsolidsphereNonM}\end{equation}
Then the external velocity in-flow is
\begin{equation}
v(r)=\sqrt{\displaystyle{\frac{K}{r}+\frac{\overline{\beta}}{r^{\alpha/2}}}},\mbox{\ \ } r > R, 
\label{eqn:nonmininflow}\end{equation} 
and the external gravitational acceleration is given by
\begin{equation}
g(r)=-\frac{\displaystyle{(1+\frac{\alpha}{2}+O(\alpha^2)\mbox{\
})}MG}{r^2}-\frac{\alpha\overline{\beta}}{4r^{1+\alpha/2}},\mbox{\ \ } r>R,
\label{eqn:externalgMin}\end{equation}
which asymptotically is dominated by the non-Newtonian second term, which essentially decreases like $1/r$.
The first term is of course  Newton's Inverse Square Law.  In this non-minimal attractor case we have a
superposition of a minimal attractor, with its strength given by the results in Sect.\ref{section:minimal},
and an independent vacuum attractor with strength $\overline{\beta}$. This happens because in the case of a
static spherically symmetric system the flow equation is linear. Of course in a physical situation these
two components would interact because the matter density would respond to the total  gravitational
acceleration. Here we have ignored this dynamical effect.

One important consequence of this form for $g(r)$ is that asymptotically Kepler's orbital laws\index{Keplerian
orbit} are violated. For circular orbits the  centripetal acceleration relation $v_O(r)=\sqrt{r g(r)}$ gives the
orbital speed to be 
\begin{equation}
v_O(r)=\left(\frac{M+M_{DM}}{r}+\frac{\alpha\beta}{4r^{\alpha/2}}\right)^{1/2},
\label{eqn:orbitalspeed}\end{equation} 
which gives an extremely flat rotation curve.  Such rotation curves are well known from  observations of
spiral galaxies.

\section[ Non-Spherical  Attractor]{ Non-Spherical Gravitational Attractors\label{section:nonspherical}}

The in-flow equation (\ref{eqn:f3extend}) also has stationary non-spherical attractors  \index{non-spherical
attractor} of the form
\begin{equation}
{\bf v}({\bf r})=\frac{{\bf
r}}{r}\left(\frac{\beta}{r^{\alpha/2}}+\frac{q}{r^\gamma}\cos(\theta)+... 
\right)^{1/2},
\label{eqn:nonmspherical}\end{equation}
where $\theta$ is the angle measured from some fixed direction, and where
\begin{equation}
\gamma=\frac{2+\alpha+\sqrt{36-4\alpha+\alpha^2}}{4}\approx 2+\frac{\alpha}{6}.
\label{eqn:gammasolution}\end{equation}
So the non-spherical term falls off quickly with distance.

\section{ Fractal Attractors\label{section:fractalattractors}}

In the early universe there would have been primordial gravitational attractors\index{primordial 
attractors} of various strengths, as defined by their $\beta$ values.  It is unknown what spectrum of $\beta$ values
would have occurred.    These would initially have all been devoid of matter agglomerations, that is they would be
`bare' attractors, because of the high temperatures.   Each such attractor represents an in-flow of space which
would have been in competition with the overall growth of space, as described previously. Each such
attractor would have a region of influence, beyond which its flow field and consequently its gravitational
field would  be cancelled by that of other  attractors. That is, the in-flow  would be confined to that
region.   Clearly attractors with larger $\beta$ values would have larger regions of influence.  Hence
space would be demarcated into a cellular form.  However within each such cellular region  there would be
smaller attractors, and within their regions, further smaller attractors.  We would then expect a fractal
cellular structure: cells within cells and so on.  This form is predicted by the emergent geometry of the 
gebit structure, as discussed in \cite{NovaBook,RC01}, and so we appear to be seeing the
linking of the bottom-up approach, from the information-theoretic ideas, with the top-down
phenomenological description of space and gravity, that has arisen from generalising the flow formalism
of both  Newtonian gravity and General Relativity.   This fractal cellular structure \index{fractal attractors} is
consistent with the in-flow equation.  Each cell would respond gravitationally to the gravitational field of the
cell in which it is effectively embedded.  Presumably attractors can merge, though this has not been analysed so
far.  Over time one would then expect that   extremely strong and spatially extended attractors would arise. 
As the universe cooled and the plasma recombined to form a neutral gas, that  gas would have been rapidly attracted by
the long range gravitational fields. Detailed studies of the dynamics of this system of fractal attractors is
required, as it is this system that determined the matter distribution of the universe, though as well we need to take
account of possible vortex systems, as discussed later.

\section{ Globular Cluster Black Holes\label{section:globular}}

\begin{figure}[h]
\vspace{8mm}
\hspace{25mm}\includegraphics[scale=0.5]{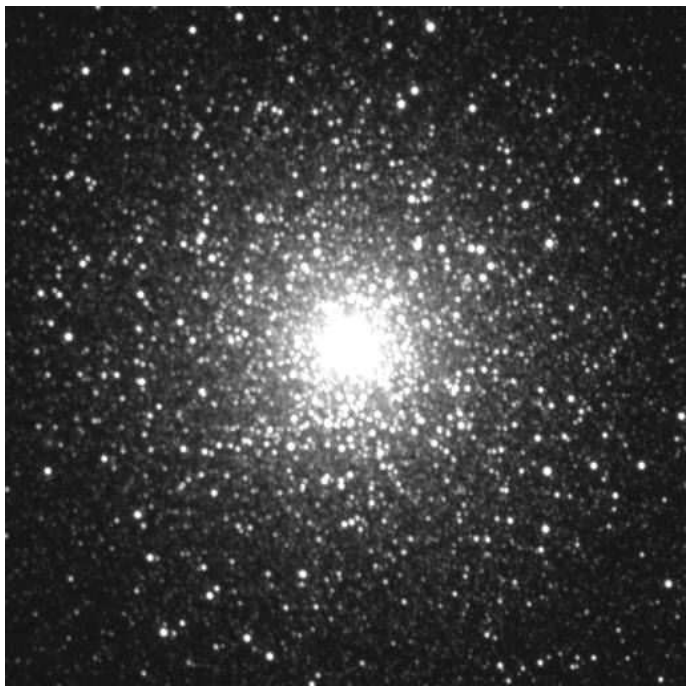}
\vspace{0mm}\caption{\small Globular cluster M15 in the constellation Pegasus, 
about 40,000 light years away, contains some 30,000 stars. M15 is one of some 150 known globular clusters
that form a halo surrounding the Milky Way. The core is tightly packed. The new theory of gravity implies that
this and other clusters have a minimal attractor, a black hole, at the centre of mass
$M_{DM}=\frac{\alpha}{2}M$. 
 }\label{fig:M15}
\end{figure}

\begin{figure}
\hspace{40mm}\includegraphics[scale=.75]{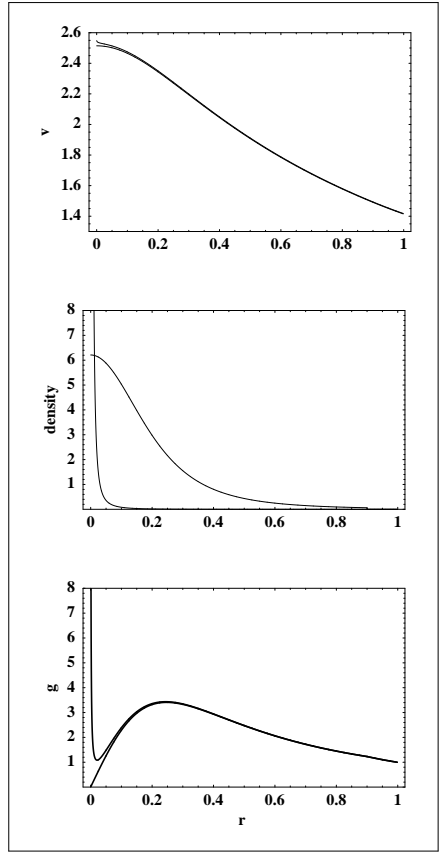}
\caption{{\small Assuming a matter density profile falling off like $1/(r^2+b^2)^2$, appropriate for a 
globular cluster,  the in-flow speed was computed from the in-flow equation, as shown in the upper plot,
which is  larger than the `Newtonian in-flow' speed near $r=0$,  as also shown in the plot. The difference becomes
very large for small $r$, but this is not shown in the plot. The matter density and the effective `dark matter'
density are shown in the middle plot. The lower plot shows the gravitational acceleration, with the strong peak at
$r=0$ caused by the induced minimal attractor. The effective mass of the attractor is given by
$M_{DM}=\alpha M/2$ to good accuracy.  A non-minimal attractor would give even larger effects
near $r=0$.   }\label{fig:GlobularPlots8}}
\end{figure}

Astronomers using the Hubble Space Telescope announced \cite{M15,G1} \index{Hubble Space Telescope (HST)}  that
they had discovered evidence for  intermediate mass black holes (IMBH),\index{intermediate mass black holes (IMBH)}
with masses from 100 to tens of thousands of solar masses.  They  believed that they had already established the
existence of stellar black holes with  masses from a few  to ten solar masses. Stellar black holes were believed to
be formed by the collapse of the cores of giant stars. But two of the Milky Way globular clusters, M15 and G1, suggested
the existence of medium sized black holes.  M15, shown in Fig.\ref{fig:M15}, is in the constellation Pegasus, while G1
is near the Andromeda Galaxy.  These clusters are some of the nearly 150 known globular clusters \index{globular
clusters} that form a halo surrounding the Milky Way. It is believed that the rising density of stars towards the
centre had resulted in a collapse of the core, leaving an IMBH.  Using the motion of stars within the clusters the mass
of the cluster and of the `black hole' were determined.  In General Relativity the mass of such a black hole would
depend on how many stars had been drawn into the black hole, and that is not predictable  without generating some
scenario for the dynamical history  of the globular cluster. However in the new theory of gravity we must have at least
a minimal gravitational attractor, whose mass is computable, and  is given to sufficient accuracy by the perturbative
result.  So the globular clusters M15 and G1  give an excellent 
opportunity to test the presence of the attractor and its effective mass.

Numerical solutions of (\ref{eqn:BHInFlowRadial}) for a  typical cluster
density profile are shown in Fig.\ref{fig:GlobularPlots8} and revealed that indeed the central  `dark
matter'  attractor has a mass  accurately given by the perturbative  result $M_{DM}/M=\alpha/2=0.00365$. 
  For M15 the mass of the central
`black hole' was found to be \cite{M15} $M_{DM}=1.7^{+2.7}_{-1.7}\times10^3$M$_\odot$, and the total
mass of M15 was determined  \cite{M15Mass}  to be $4.9\times10^5$M$_\odot$.  Then  these results together
give   $M_{DM}/M=0.0035^{+0.011 }_{-0.0035}$  which is in
excellent agreement with the above prediction. For G1 we have \cite{G1} 
$M_{DM}=2.0^{+ 1.4 }_{-0.8 }\times 10^4$M$_\odot$, and $M=(7-17)\times10^6$M$_\odot$. These values give
$M_{MD}/M= 0.0006-0.0049$, which is also consistent with the above $\alpha/2$ prediction. 

However there is one complication in this analysis. The determination of the `black hole' mass followed
from stochastic modelling of the motion of the inner stars, which was  compared with the motion of those
stars as revealed by the HST. In that modelling the gravitational acceleration caused by the `black hole'
would have been described by Newton's inverse square law form: $g(r) \sim 1/r^2$. However the attractor
produces a gravitational acceleration of the form: $g(r) \sim 1/r^{1+\alpha/2}$.  Hence to test the
attractor explanation  it is necessary for the stochastic modelling to be repeated using this modified
force law.  Nevertheless that the attractor explanation gives masses consistent with the observations and
modelling is very encouraging.  The attractor explanation is independent of the dynamical history of the
globular cluster.  Observations of other clusters should confirm that they all have the same mass
ratio $M_{DM}/M=\alpha/2$.  Of course there is an event horizon associated with the attractor, and to that
extent we can continue to describe the attractor as a `black hole', though one very different from that of
General Relativity.   
 
\section[ Galactic Rotation Curves and Gravitational Attractors]{  Galactic Rotation Curves and Gravitational
\newline \mbox{\  }Attractors\label{section:spiral}}

Consider the case of  a spiral galaxy with a non-spherical rotating matter
distribution. 
For spiral galaxies the new theory of gravity implies that their central black holes are large  non-minimal
primordial attractors.  Then the non-inverse square law acceleration of such an attractor, in (\ref{eqn:externalgMin}),
would have caused a vary large in-fall speed for the surrounding matter, unlike the Newtonian-like gravitational in-fall
which would result from the matter alone without a primordial attractor. Such a large in-fall speed would almost
certainly result in a large angular momentum for the matter, resulting in the rotating flat disk so charactersitic of
spiral galaxies. Then by the same argument we see that non-rotating elliptical galaxies are formed by a
gravitational in-fall mechanism that does not have a central primordial black hole, or at least only a very small
one. Hence elliptical galaxies should not display the  extreme `dark matter' effect that follows from the
presence of a central non-minimal attractor, as is apparent now from a recent analysis of elliptical galaxies 
\cite{Elliptical}. As well the region of influence of a primordial attractor will determine the total amount of
matter that it can attract to form a spiral galaxy, and so there will be a relationship between the total `dark
matter' content of a spiral galaxy and its luminosity.  On the other hand for elliptical galaxies the central
attractors are simply a consequence of the matter induced in-flow, resulting in a minimal attractor, so their
central black hole mass should be related to their total matter content by the same relationship as for the
globular clusters, with some correction for their non-sphericity, which in general also permits more than one
such black hole. 

For the case of spiral galaxies we need to use  numerical techniques, but beyond a sufficiently large
distance  the in-flow,  due then mainly to the primordial attractor, will have spherical symmetry, and in that region we
may use (\ref{eqn:BHInFlowRadial})  and (\ref{eqn:BHdm1}) with
$\rho(r)=0$.  Then  as already analysed the in-flow has the form, on re-parametrising (\ref{eqn:nonmininflow}),
\begin{equation}
v(r) = {\overline K}\left(\frac{1}{r}+\frac{1}{R_S}\left(\frac{R_S}{r}  \right)^{\displaystyle{\frac{\alpha}{2}}} 
\right)^{1/2},
\label{eqn:vexact}\end{equation}
where ${\overline K}$ and $R_S$ are arbitrary constants in the $\rho=0$ region, but where the value
of ${\overline K}$ is determined by matching to the solution in the matter region. Here $R_S$ characterises the
length scale of the non-perturbative non-minimal attractor part of this expression,  and ${\overline K}$ depends on
$\alpha$ and $G$ and details of the matter distribution.  The galactic circular orbital velocities of stars etc may
be used to observe this process in a spiral galaxy and  from (\ref{eqn:ga}) and (\ref{eqn:vexact}) we obtain a
replacement for  the Newtonian  `inverse square law',
\begin{equation}
g(r)=\frac{{\overline K}^2}{2} \left( \frac{1}{r^2}+\frac{\alpha}{2rR_S}\left(\frac{R_S}{r}\right)
^{\displaystyle{\frac{\alpha}{2}}} 
\right),
\label{eqn:gNewl}\end{equation}
in the asymptotic limit.     From  (\ref{eqn:gNewl}) the centripetal
acceleration  relation
$v_O(r)=\sqrt{rg(r)}$  gives  a `universal rotation curve'
\begin{equation}
v_O(r)=\frac{{\overline K}}{2} \left( \frac{1}{r}+\frac{\alpha}{2R_S}\left(\frac{R_S}{r}\right)
^{\displaystyle{\frac{\alpha}{2}}} 
\right)^{1/2}.
\label{eqn:vorbital}\end{equation}
\begin{figure}[t]
\hspace{30mm}\includegraphics[scale=1.1]{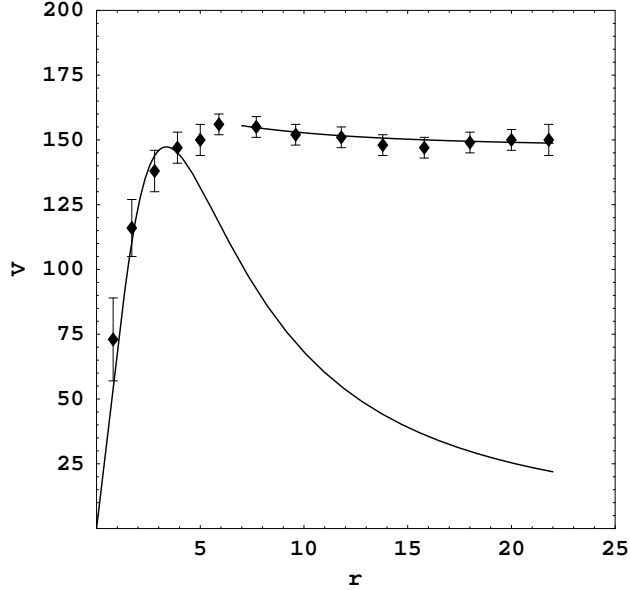}
\caption{{\small Rotation-velocity curve plot for the spiral galaxy NGC3198, with $v$ in km/s, and $r$ in kpc/h.
Complete curve is rotation curve expected from Newtonian theory of gravity or the General Theory of Relativity for an
exponential disk, which decreases asymptotically like $1/\surd r$. The incomplete  curve shows the asymptotic form 
from (\ref{eqn:vorbital}).}  
\label{fig:NGC3198}}\end{figure}
\begin{figure}[t]
\hspace{30mm}\includegraphics[scale=1.1]{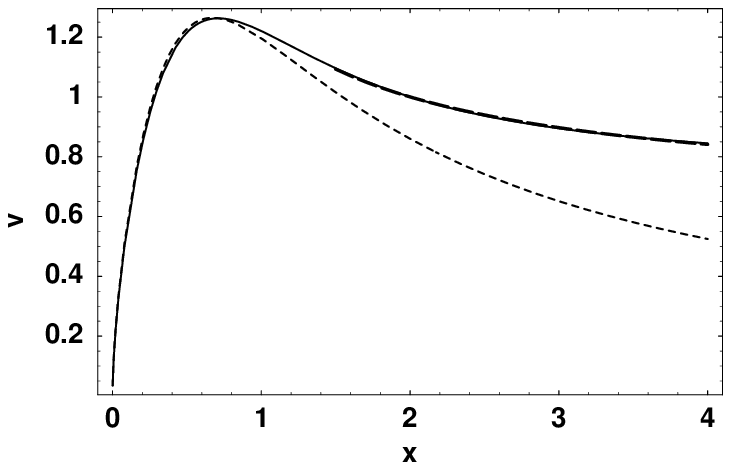}
\caption{{\small Spiral galaxy rotation velocity curve plots, with  $x=r/R_{opt}$.  Solid line is the Universal
Rotation Curve (URC) for  luminosity
$L/L_*=3$, using the URC in  (\ref{eqn:URC}), Ref.\cite{URC}. Short dashes line is URC with only the matter
exponential-disk contribution, and re-fitted to the full URC at low x. Long dashes  line, which essentially
overlaps the solid line for $x > 1.5$, is the form  in (\ref{eqn:vorbital}) for $\alpha=1/137$ and
$R_S=0.01R_{opt}$.}  
\label{fig:URCplots}}\end{figure}
Because of the $\alpha$ dependent part this rotation-velocity curve \index{rotation velocity curve}  falls off
extremely slowly with $r$, as is indeed observed for spiral galaxies. Of course it was the inability of the 
Newtonian  and Einsteinian gravity theories to explain these observations that led to the  notion of `dark
matter'\index{dark matter effect}. It is possible to illustrate the form in (\ref{eqn:vorbital}) by comparing it
with rotation curves of spiral galaxies\index{spiral galaxies}.  Fig.\ref{fig:NGC3198} shows the rotation curve
for the spiral galaxy NGC3198. Persic,  Salucci and   Stel
\cite{URC} analysed 
 some 1100 optical and radio rotation curves, and demonstrated that they are describable by the  empirical 
universal rotation curve (URC)\index{universal rotation curve (URC)}
\begin{eqnarray}
v_O(x)
&=&v(R_{opt})\left[\left(0.72+0.44\mbox{Log}\frac{L}{L_*}\right)\frac{1.97x^{1.22}}{(x^2+0.78^2)^{1.43}}\right.\nonumber \\
 & &\mbox{\ \ \ \ \ \ \ \ \ \ \ \ }\left.+1.6\mbox{e}^{-0.4(L/L_*)}
\frac{x^2}{x^2+1.5^2(\frac{L}{L_*})^{0.4}}\right]^{1/2},
\label{eqn:URC}\end{eqnarray}
where $x=r/R_{opt}$, and where $R_{opt}$ is the optical radius, or $85\%$ matter limit.  The first term is the
Newtonian contribution from an exponential matter disk, and  the 2nd term is the `dark matter' contribution. 
This two-term form also arises  from the in-flow theory, as follows from (\ref{eqn:veqnagain}). 
   The form in (\ref{eqn:vorbital}) with
$\alpha=1/137$  fits, for example, the high luminosity URC, for a suitable value of $R_S$, which depends on the luminosity,
as shown by one example in Fig.\ref{fig:URCplots}.  For low luminosity data the observations do not appear to extend far enough
to reveal the asymptotic form of the rotation curve, predicted by (\ref{eqn:vorbital}). 

  As already noted in
(\ref{eqn:DMsolidsphereNonM}) the effective `dark matter' density for these non-minimal attractors falls  off,
essentially, like
$1/r^2$, as astronomers had deduced from observations of the rotation curves of spiral galaxies. This leads to the
following expression for the total `dark matter' within radius $R$
\begin{equation}
M_{r<R}=4\pi\int_0^R\rho_{DM}(r)r^2dr=\frac{\alpha\beta R^{1-\alpha/2}}{4G},
\label{eqn:totalM}\end{equation}
which increases almost linearly with $R$. It is this expression that explains the observations that the total amount
of `dark matter' exceeds the real amount of matter, as revealed by its luminosity, by often an order of magnitude.

\section{  Stellar Structure\label{section:stellarstructure}}

The structure of stars is very much based on the assumption that the Newtonian theory of
gravity is sufficiently accurate. This leads to the Solar Standard Model (SSM) in the case of the sun. However
the new theory of gravity predicts at least a minimal gravitational attractor\index{minimal stellar
attractor} at the centres of  stars, with an associated event horizon.  This quantum-foam in-flow singularity
causes the gravitational acceleration to be very different to that from the Newtonian theory, as already
illustrated by a number of  cases.   Bahcall and Davis started an
exploration of the sun by means of neutrinos\index{solar neutrino anomaly} \cite{Bahcall,Davis}, with that work
resulting in the solar neutrino anomaly, namely that all experiments, exploring different portions of the solar
neutrino spectrum, reported a flux less than that predicted.   The solar  neutrino  flux is determined by the
physical and chemical properties of the sun, such as density, temperature, composition and so on.  The current
interpretation of the solar neutrino anomaly is in terms of the neutrino masses and mixing  leading to oscillations
of $\nu_e$ into active ($\nu_\mu$ and/or
$\nu_\tau$) or sterile neutrino, $\nu_s$.  

Because of the gravitational attractor effect one of the key equations in the SSM, the equation for
hydrostatic equilibrium,
\begin{equation}
\frac{dP(r)}{dr}=-\rho(r)g(r)=-\frac{G\rho(r)m(r)}{r^2},
\label{eqn:hydro}\end{equation}
where $P(r)$ is the hydrostatic pressure, $\rho(r)$ is the matter density and $m(r)$ is the matter within
radius $r$, must be changed to
\begin{equation}
\frac{dP(r)}{dr}=-\frac{G\rho(r)m(r)}{r^2}-\frac{2\pi\alpha\rho(r) G}{r^2}\int_0^r\left(\int_s^R
s^\prime\rho(s^\prime) ds^\prime\right) ds,
\label{eqn:hydronew}\end{equation}
from (\ref{eqn:BHISL2}) to $O(\alpha)$ terms. Towards the centre of the star this equation is dominated by the
$\alpha$ dependent term. We can illustrate this for the simple case of uniform density.  In this case
(\ref{eqn:hydro})  gives
\begin{equation}
P(r)=\frac{2\pi G \rho^2(R^2-r^2)}{3},
\label{eqn:oldstar}\end{equation}
where $R$ is the radius of the star, and where $P(R)=0$, giving a finite pressure at the centre $r=0$.
However (\ref{eqn:hydronew}) becomes
\begin{equation}
\frac{dP(r)}{dr}=-\frac{4}{3}\pi G \rho^2 r- \pi \alpha \rho^2 G(\frac{R^2}{r}-\frac{r}{3}),
\label{newstar}\end{equation}
with solution
\begin{equation}
P(r)=\frac{2\pi(1+\alpha/4) G \rho^2 (R^2-r^2)}{3}+\pi\alpha G \rho^2
R^2\mbox{ln}(\frac{R}{r}),
\label{eqn:newpressure}\end{equation}
which reveals a logarithmic pressure singularity at the centre.  In a more realistic modelling
this  effect would probably be even more pronounced, as the density would increase there as a consequence of such a
pressure increase.  Existing   stellar structure codes need modification in order for this effect to be
explored, and for any signature of the effect on the neutrino flux revealed.  Because there is an event
horizon at the centre of stars, essentially a black hole effect, though one very different from that of
general relativity,  an additional source of energy, and hence heating is available, namely the radiation
from matter falling into this black hole.  This would, even by itself,  increase the temperature of the
central region of stars.

Radiometric data from the Pioneer 10/11\index{Pioneer 10//11 anomaly}, Galileo, and Ulysses spacecraft indicated an
apparent anomalous, constant, acceleration acting on the spacecraft with a magnitude of $\sim 8.5\times10^{-8}$ cm/s$^2$ 
directed towards the sun \cite{Pioneer}. For Pioneer 10/11 the acceleration was 
$ 8.56\times10^{-8}$ cm/s$^2$  at 30 AU, while at 60 AU it was $ 8.09\times10^{-8}$ cm/s$^2$.  If there was a
non-minimal gravitational attractor associated with the sun, we would expect an anomalous acceleration directed
towards the sun, but decreasing like $1/r$.  However the Pioneer 10/11 data does not indicate any such 
decrease, and so we conclude that (i) there is no evidence yet for such a non-minimal  attractor, and (ii) that
the new theory of gravity does not offer an explanation for this anomalous acceleration. 

\section[ Quantum Gravity Experiments]{ Laboratory Quantum Gravity Experiments}\label{section:laboratory}
Quantum gravity\index{quantum gravity} effects are really just  quantum-foam in-flow effects. As discussed in
Sect.\ref{section:measurementsofG} such effects have been manifest in ongoing attempts to measure $G$ over the last
60 years. There they showed up as
$O(\alpha)$ unexplained systematic effects.   The new theory of gravity has two fundamental constants $\alpha$ and $G$,
and clearly one cannot measure one of these alone.   These $G$ measurement experiments basically measure the force
between two test masses as a function of separation distance. Using the linearity of the Newtonian theory  of gravity
the  computation of these forces involves a vector sum of the forces between the individual mass points in the different
test masses.  However in the new theory there is a non-linearity  whose magnitude is determined by $\alpha$.

\begin{figure}
\hspace{0mm}\includegraphics[scale=1.6]{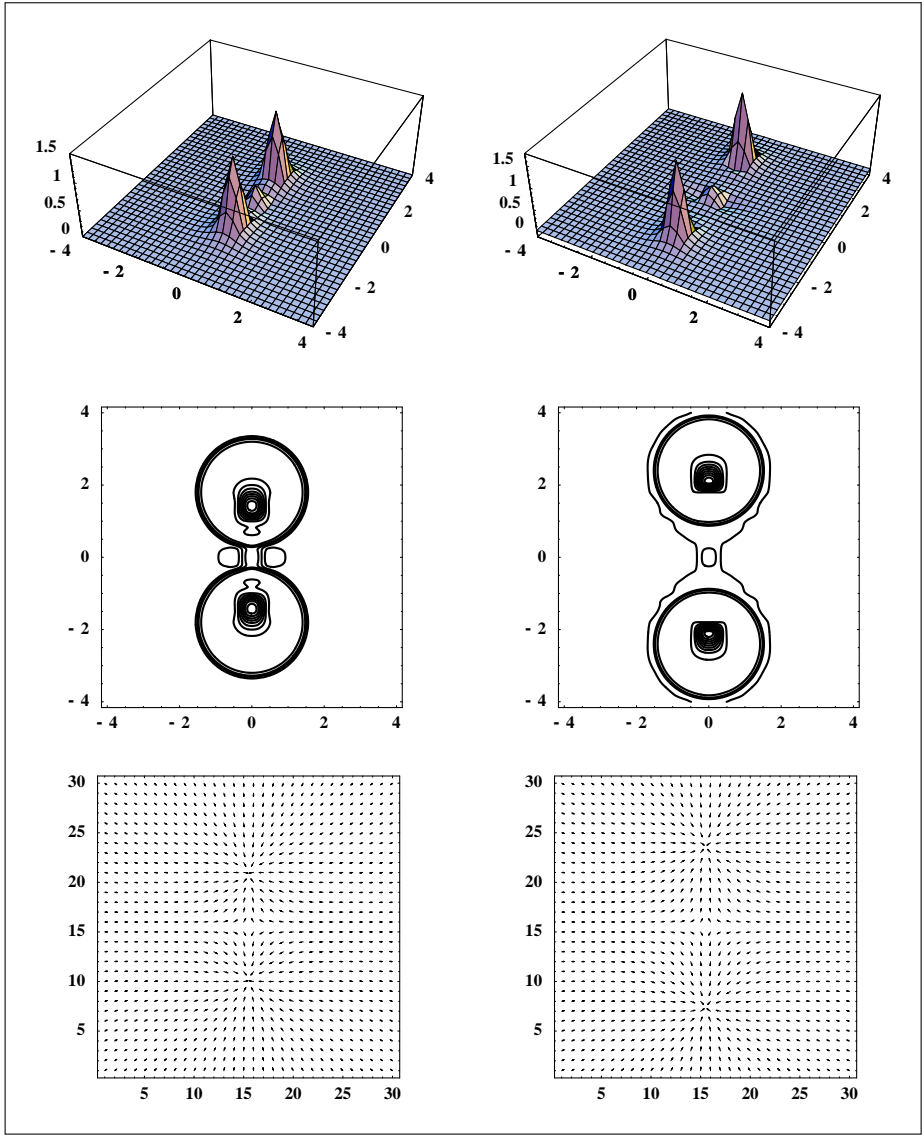}
\caption{{\small Cavendish experiment with two spheres, each  of radius 1.5,  and with uniform density. On the left the
separation of the centres is 3.6, while on the right the separation is 4.8.  The upper plots
show the `dark matter' density distributions of the gravitational attractors, as also shown in the middle
contour plots. The contour plots clearly reveal the `polarisation' effect  of the `dark matter' density, which is greater
for smaller separations. In each case the ratio of the total `dark matter' to the total mass is  0.0045. The lower plots
show the in-flow velocity fields. These are quantum gravity effects that are detectable in laboratory experiments. 
}\label{fig:Twosphereplots}}
\end{figure}

Assuming
a stationary in-flow the velocity field is given by
\begin{equation}\label{eqn:QG}
|{\bf v}({\bf r})|^2=\frac{2}{4\pi}\int d^3
r^\prime\frac{C({\bf v}({\bf r}^\prime))}{|{\bf r}-{\bf r}^\prime|}-2\Phi({\bf r}).
\end{equation} 
with the {\it ansatz} that the direction of ${\bf v}(\bf r)$ is the same as the direction of ${\bf g}(\bf r)$, where this
gravitational acceleration is given by
\begin{equation}
{\bf g}{(\bf r)}=\frac{1}{2}{\bf \nabla}({\bf v}^2(\bf r)).
\label{eqn:QG2}\end{equation}

It is easier to see the effects of the non-linearity of the `dark matter'  term by computing and displaying this
matter density in those cases relevant to a Cavendish-type laboratory experiment in which both $\alpha$ and $G$ are
measured together. In Fig.\ref{fig:Twosphereplots} \index{quantum gravity experiments} is shown the `dark
matter'\index{Cavendish!dark matter effect} density for two spheres for the cases of two separation distances. Here we
have ignored the inhomogeneity of the in-flow of the earth. When the spheres are well separated the `dark matter'
effect occurs at the centre of each sphere, but as they are brought together the non-linearity of the effect
causes the `dark matter' density to become essentially polarised, this `dark matter' polarisation is evident in
Fig.\ref{fig:Twosphereplots}, that is, in each mass it moves away from the centre, and there is also
a small region of `dark matter' that forms outside and between the two spheres.  Because of this non-linear polarisation
effect\index{attractors and polarisation effects} the net force between the two spheres is not  describable by the  
Newtonian theory.  However the leading effect is that the `dark matter' mass depends on the geometry of the masses
used. The deviations from the Newtonian theory then permit the experimental determination of
$\alpha$ from such an experiments\index{fine structure constant ($\alpha$)}.  This effect is a key prediction and
provides a critical test of the new theory.  An analysis of the data requires numerical solutions of the flow
fields for each configuration of the masses in such an experiment.    The magnitude  of the `dark matter' mass 
is $\approx \alpha/2$, i.e. 0.3\%, but the main observable effect is the dependence of this mass on the geometry
of the objects, which experiment suggests is at the 0.1 \% level. 

\newpage

\section{ Conclusions}
We have seen that the solar system was too special to have revealed key aspects of gravity. These only become 
evident when Newtonian gravity is re-formulated in terms of a velocity in-flow field.  A  generalisation of that
formalism leads to an explanation of the so-called `Dark Matter' effect.  The most significant aspect of this work
is the discovery that the magnitude of the `dark matter' spatial self-interaction dynamics is determined by the fine
structure constant $\alpha$, while Newton's gravitational constant $G$ only determines the interaction of space with
matter. It has been shown how the values of $\alpha$ and $G$ together may be determined in Cavendish-type
laboratory experiments.  Also significant is the discovery that this new theory of gravity leads to gravitational
attractors - a new form of black hole, and whose properties are determined by the value of $\alpha$ and not $G$.
As a consequence we saw how these attractors explain the observed globular cluster black hole masses, and similarly
the necessity for black holes in galaxies.  We also saw that  primordial non-minimal
black holes explain both the  origin and nature of spiral galaxies, particularly their rotational behaviour.
This new theory of gravity has its origins in the information-theoretic {\it Process Physics}, which offers as well
an explanation and unification of the quantum nature of space and matter, and their classicalisation \cite{NovaBook}.
That work implies that the occurrence of $\alpha$ in the new classical description of gravity is indicative of the
underlying quantum processes  of space, that space is a complex quantum-information theoretic system, and at that
level it is not geometric.  The results herein imply that we have the first evidence of  quantum aspects of
gravity.

\end{document}